\documentclass[aps,twocolumn,prb,10pt, floatfix,longbibliography,final]{revtex4-2}
\usepackage[caption=false]{subfig}
\usepackage{graphicx, amsmath, bm, amsfonts, amssymb, color}
\usepackage[%
  pdftex,
  breaklinks=true,
  colorlinks=true,
  linkcolor=blue,
  urlcolor=blue,
  citecolor=blue
]{hyperref}
\usepackage{verbatim}
\usepackage{xcolor}
\usepackage{booktabs}

\graphicspath{ {./images/} }

\newcommand{\beq}[0]{\begin{equation}}
\newcommand{\eeq}[0]{\end{equation}}
\newcommand{\beqa}[0]{\begin{eqnarray}}
\newcommand{\eeqa}[0]{\end{eqnarray}}

\usepackage{enumitem}

\newcommand{\Addition}[1]{\color{black} #1 \color{black}}

\begin{document}

\title{
Nonequilibrium Quasiparticles in Superconducting Circuits: Energy Relaxation, Charge and Flux Noise
}
\author{Jos\'{e} Alberto Nava Aquino}
\author{Rog\'{e}rio de Sousa}
\affiliation{Department of Physics and Astronomy, University of Victoria, Victoria, British Columbia V8W 2Y2, Canada}
\affiliation{Centre for Advanced Materials and Related Technology, University of Victoria, Victoria, British Columbia V8W 2Y2, Canada}

\date{\today}

\begin{abstract}
The quasiparticle density observed in low-temperature superconducting circuits is several orders of magnitude larger than the value expected at thermal equilibrium. The tunneling of this excess of quasiparticles across Josephson junctions is recognized as one of the main loss and decoherence 
mechanisms in superconducting qubits. 
%
Here we present a unified impedance theory that accounts for quasiparticle energy loss in circuit regions both 
far and near (across) junctions.
%
Our theory leverages the recent experimental demonstration that the excess quasiparticles are in \emph{quasiequilibrium} [T. Connolly {\it et al.}, \href{https://doi.org/10.1103/PhysRevLett.132.217001}{Phys. Rev. Lett. {\bf 132}, 217001 (2024)}] and uses a generalized fluctuation-dissipation theorem to predict the amount of charge and flux noise generated by them. 
We compute the resulting energy relaxation time $T_1$ in transmon qubits with and without junction asymmetric gap engineering, and show that quasiparticles residing away from junctions can play a dominant role in the former case. They also may provide an upper limit for resonator quality factors if the density of amorphous two-level systems is reduced. 
%
%
In addition, we show that charge noise from quasiparticles leads to flux noise that is  logarithmic-in-frequency, giving rise to a ``nearly white" contribution that is comparable to the flux noise observed in flux qubits. This contrasts with amorphous two-level systems, whose associated flux noise is shown to be superOhmic. We discuss how this quasiparticle flux noise can limit $T_2^{*}$ coherence times in flux-tunable qubits. The final conclusion is that asymmetric gap engineering can greatly reduce noise and increase coherence times in superconducting qubits. 
%
%
\end{abstract}

\maketitle

\section{Introduction\label{sec:introduction}}

Superconducting (SC) qubits represent a promising pathway toward scalable quantum computing, leveraging the coherence of macroscopic quantum states \cite{Clarke2008}. A pivotal factor in their operational efficacy is maintaining quantum coherence, a challenge compounded by various decoherence mechanisms \cite{Martinis2005, Koch2007, Sendelbach2008, Gao2008, deSousa2014, Nava2022}. Among these, quasiparticles (QPs), excitations resulting 
from broken Cooper pairs, emerge as a concern. The presence of QPs in superconductors is known to give rise to surface resistance and Ohmic loss \cite{Mattis-Bardeen1958}.  
Their impact on superconducting (SC) qubits is believed to be the greatest when they tunnel across a Josephson junction (JJ), leading to energy relaxation and dephasing, thereby limiting qubit performance \cite{Lutchyn2006, Martinis2009, Catelani2011, CatelaniPRL2011}.

Several experiments show that a large population of nonequilibrium QPs (resident QPs) remain even at low temperatures ($k_BT\ll \Delta$), in spite of the thermal QP population being exponentially small ($\propto e^{-\Delta/k_BT}$, where $\Delta$ is the SC energy gap) \cite{Aumentado2004, Serniak2018, Liu2024}.
These nonequilibrium populations \Addition{were shown to arise from external perturbations, such as ionizing radiation \cite{Vepsalainen2020} and stray infrared photons \cite{Diamond2022, Du2022}. It was also pointed out that QPs may be produced from the high-frequency components of the charge noise from e.g. two-level systems in the materials surrounding the superconductor \cite{Alase2025}.}

An additional 
unknown is
the energy distribution for the resident QPs. A recent experiment provided evidence of \emph{quasiequilibrium}, where QPs are in thermal equilibrium with the surrounding phonon bath despite having an out-of-equilibrium density. Therefore, even though the QPs arise from high-energy sources, rapid inelastic processes mediated by phonons restore them to a thermal-like energy distribution \cite{Connolly2024}. 

Current designs of SC circuits engineer junction asymmetries in order to prevent QP-tunneling across the circuit's JJs, greatly reducing the impact of the QP-tunneling mechanism \cite{Yamamoto2006, Marchegiani2022, Connolly2024, McEwen2024}. 

Here we present a unified impedance theory for Ohmic loss generated by QPs in SC circuits. 
The theory treats the Ohmic loss due to QP tunneling across junctions on equal footing as the Ohmic loss within SC wire regions away from junctions. We calculate the charge and flux noise spectral densities that emerge from this generalized Ohmic loss, and use it to predict energy relaxation times $T_1$  and coherence times $T_{2}^{*}$ for superconducting qubits. 
Among other results we show that QPs away from junctions may limit the energy relaxation time $T_1$ of gap-engineered transmon qubits. 

%
%


We also present explicit numerical calculations for capacitively-coupled waveguide resonators (CPW resonators) to show that high densities of resident QPs within the wires limits the quality factor of CPWs to less than $10^{7}$. 

%

Additionally, we show that charge noise from QP Ohmic loss both near and far from junctions generates flux fluctuations due to the self-inductance of the wires, giving rise to a flux noise background that is logarithmic in frequency. The magnitude of the predicted ``nearly white flux noise background'' is found to be comparable to values observed in flux qubit experiments \cite{Quintana2017, Luthi2018}. 

This QP flux noise is large at low frequencies and may impose significant limitations to the coherence times $T_{2}$ and $T_{2}^{*}$ of flux-tunable qubits such as the split-junction transmon. Flux tunable qubits are considered to be the solution to the problem of frequency crowding in quantum computers with a large number of qubits, but their frequency tunability makes them sensitive to intrinsic flux fluctuations \cite{Hutchings2017}. This is true in spite of the fact that all transmons (including the split-junction ones) are designed to be insensitive to pure charge noise \cite{Koch2007}. 

We show that this mechanism leads to logarithmic-in-time qubit coherence decay that limits $T_{2}^{*}$ in flux-tunable qubits such as the split-junction transmon.

%

\section{Qubit energy relaxation rate and resonator quality factor from charge and flux noise\label{sec:relaxation_noise}}

The rate for energy equilibration between a qubit and a charge noise environment follows from the qubit-environment coupling ${\cal H}_{{\rm int}}=\frac{\partial {\cal H}}{\partial Q}\delta Q$. Here ${\cal H}$ is the qubit Hamiltonian, with $Q$ and $\frac{\partial {\cal H}}{\partial Q}$ representing the qubit's charge and voltage operators, respectively. The operator $\delta Q$ describes charge fluctuations in the environment. From Fermi's golden rule we get \cite{Schoelkopf2003, deSousa2009} 
\begin{equation}
    \frac{1}{T_1}= \frac{1}{\hbar^2}\left|\langle 1\right|\frac{\partial {\cal H}}{\partial Q}\left|0\rangle\right|^{2}\left[\tilde{S}_Q(\Omega)+\tilde{S}_Q(-\Omega)\right],
\label{T1gen}
\end{equation}
where $|0\rangle$ and $|1\rangle$ are the qubit's lowest energy eigenstates with energy difference $E_1-E_0\equiv \hbar\Omega$. 
The impact of the environment is encoded in the charge noise spectral density,
\begin{equation}
    \tilde{S}_Q(\omega)=\int_{-\infty}^{\infty} dt e^{i\omega t}\left\langle \delta Q(t)\delta Q(0)\right\rangle_T,
\end{equation}
where 
$\langle \cdot\rangle_T$ is thermal average. 
Here $\omega$ denotes an arbitrary frequency, in order to distinguish from the resonant frequency $\Omega$ for a qubit or resonator. 
While Eq.~(\ref{T1gen}) shows that only noise at $\omega=\Omega$ contributes to $1/T_1$, off-resonant contributions at $\omega\neq \Omega$ give rise to phase relaxation processes that will contribute to $1/T_{2}^{*}$ and $1/T_{2}$ when the qubit's energy splitting $\hbar\Omega$ is sensitive to charge or flux fluctuations \cite{deSousa2009}.

As an application of Eq.~(\ref{T1gen}), assume the qubit is approximately described by a \emph{simple harmonic LC-resonator},
\begin{equation}
    {\cal H}=\frac{Q^2}{2C}+\frac{\Phi^2}{2L}, 
\end{equation}
where $C$ and $L$ are the capacitance and inductance, respectively, and $\Phi$ is the flux operator, satisfying $\left[Q,\Phi\right]=i\hbar$. In this case Eq.~(\ref{T1gen}) becomes
\begin{equation}
    \frac{1}{T_1}=\frac{\Omega}{2\hbar C}\left[\tilde{S}_Q(\Omega)+\tilde{S}_Q(-\Omega)\right], 
    \label{T1resonator}
\end{equation}
with $\Omega=1/\sqrt{LC}$. 
In Appendix~\ref{sec:GDFT} we show that the usual fluctuation-dissipation theorem remains valid when the environment satisfies the ``quasithermal law'' described below. The fluctuation-dissipation theorem relates charge noise to the linear response charge susceptibility $\tilde{\chi}_Q(\omega)=\langle \delta \tilde{Q}(\omega)\rangle/\delta \tilde{V}(\omega)$,
\begin{equation}
    \tilde{S}_Q(\omega) = 2\hbar\;{\rm Im}\left\{\tilde{\chi}_Q(\omega)\right\}\left[ n_B(\omega) + 1 \right],
    \label{SqFD}
\end{equation}
where $n_B(\omega) = 1/(e^{\hbar \omega/k_BT}-1)$ is the Bose-Einstein distribution. Inserting this into Eq.~(\ref{T1resonator}) leads to
\begin{equation}
    \frac{1}{T_1}= \frac{\Omega}{C}{\rm Im}\left\{\tilde{\chi}_Q(\Omega)\right\}\coth{\left(\frac{\hbar\Omega}{2k_BT}\right)}. 
    \label{T1chi}
\end{equation}
At low temperatures $k_BT\ll \hbar\Omega$ the $\coth$ can be approximated by $1$ leading to $1/T_1=\Omega/Q$ where $Q$ is the resonator quality factor \cite{noteA},
\begin{equation}
    \frac{1}{Q}=\frac{{\rm Im}\left\{\tilde{\chi}_Q(\Omega)\right\}}{C}\equiv\langle \tan{(\delta)}\rangle =\sum_i p_i \tan{(\delta_i)}.
\label{Qtan}
\end{equation}
Here $\langle \tan{(\delta)}\rangle$ is the loss tangent averaged over regions $i$ of the device according to the participation ratio $p_i$, which is defined as the electrical energy of region $i$ divided by the total device electromagnetic energy. 

We may also evaluate Eq.~(\ref{T1gen}) for the transmon qubit, which is a slightly anharmonic $LC$ resonator with a Josephson junction playing the role of the inductor \cite{Koch2007transmon}. The transmon is described by the Hamiltonian
\begin{equation}
    {\cal H}=\frac{E_C}{e^2}Q^2-E_J \cos\left(2\pi \frac{\Phi}{\Phi_0}\right), \label{eq:Hamiltonian_transmon}
\end{equation} 

where $E_J=\frac{\Phi_0 I_c}{2\pi}$ and $E_C=e^2/(2C)$ are the Josephson and charge energies associated to the junction, respectively ($I_c$ is the critical current and $\Phi_0=\frac{h}{2|e|}$ is the flux quantum). The phase difference across the junction is $\varphi=2\pi \frac{\Phi}{\Phi_0}$.
The $1/T_1$ is identical to Eq.~(\ref{T1resonator}) with the qubit frequency given by
\begin{equation}
    \hbar\Omega \approx \sqrt{8E_JE_C}=\frac{\hbar}{\sqrt{L_J C}},
\end{equation}
provided that $E_J\gg E_C$.
In the last equality we introduced the Josephson inductance $L_J=\left(\frac{\Phi_0}{2\pi}\right)^{2}\frac{1}{E_J}=\frac{\Phi_0}{2\pi I_c}$. Hence, the transmon can also be described by the resonator quality factor Eq.~(\ref{Qtan}). 

We now move on to describe what is often believed to be the main contribution to resonator and transmon energy relaxation rate. TLSs defects are one of the main sources of charge noise in SC circuits \cite{Wang2015, Gorgichuk2023}. Inserting the definition of loss tangent to Eq.~(\ref{SqFD}) we get the charge noise due to TLSs, 
\beq \label{eq:SQcap}
    \tilde{S}_Q^{\rm TLS}(\omega) = 2 \hbar C \langle\tan{(\delta_{TLS})}\rangle \left[ n_B(\omega) + 1 \right],
\eeq
with the TLS loss tangent averaged over surface (S) and bulk (B) dielectric regions of the device. In the low power regime this is equal to 
\begin{eqnarray}
    \langle\tan{(\delta_{TLS})}\rangle&=&\left[p_S\tan{(\delta_{TLS,0}^{S})} +p_B\tan{(\delta_{TLS,0}^{B})}\right]\nonumber\\ &&\times\tanh{\left(\hbar\omega/2k_BT\right)},
    \label{avgtandeltaTLS}
\end{eqnarray}
with 
typical amplitudes $\tan{(\delta_{TLS,0}^{S})}=10^{-3}$ and 
$\tan{(\delta_{TLS,0}^{B})}=10^{-6}$ for TLSs located in the bulk and surface regions, respectively \cite{Wang2015, Gorgichuk2023}. 


There is an alternative way of expressing Eq.~(\ref{T1gen}) that warrants discussion. The presence of charge fluctuation implies associated flux fluctuation because $\Phi=L\dot{Q}$, where $L$ is the SC wire self inductance and $\dot{Q}$ is the associated current. As a result $\delta \tilde{\Phi}= -i\omega L \delta \tilde{Q}$, implying the flux noise spectral density 
\begin{equation}
    \tilde{S}_{\Phi}(\omega) =  (L \omega)^2\tilde{S}_Q(\omega)
    \label{eq:SPhigen}
\end{equation}
%
is always associated to charge noise. In this case Eq.~(\ref{T1gen}) can be rewritten as 
\begin{equation}
        \frac{1}{T_1}= \frac{1}{\hbar^2}\left|\langle 1\right|\frac{\partial {\cal H}}{\partial \Phi}\left|0\rangle\right|^{2}\left[\tilde{S}_\Phi(\Omega)+\tilde{S}_\Phi(-\Omega)\right].  
        \label{T1genPhi}
\end{equation}
It must be emphasized that if the origin of flux noise is charge fluctuation, then 
Eqs.~(\ref{T1gen})~and~(\ref{T1genPhi}) describe the same mechanism and should not be added together. However, Eq.~(\ref{T1genPhi}) can be used to account for additional mechanisms not associated to charge fluctuation, e.g. flux noise due to spin-impurity fluctuation \cite{Nava2022}. 

\section{Charge Susceptibility, impedance and conductivity from Quasiparticles \label{sec:qp_impedance}}

In this section we propose analytical approximations for the charge susceptibility and impedance
of a SC circuit due to resident quasiparticles. 


\subsection{Two-fluid model for superconducting wires \label{subsec:wire}}

The linear response in a general circuit due to a small perturbation of the voltage $V$ is $\delta \tilde{V} = Z \langle \delta \tilde{I} \rangle= -i \omega Z \langle \delta \tilde{Q} \rangle$, where $Z$ is a complex impedance.
The charge susceptibility in a circuit is thus $\tilde{\chi}_Q(\omega)  = \frac{\langle \delta \tilde{Q} \rangle}{\delta \tilde{V}} = - \frac{1}{i \omega Z}$, and its imaginary part is
\beq
    {\rm Im} \left\{ \tilde{\chi}_Q(\omega) \right\} 
= \frac{1}{\omega}{\rm Re} \left\{ \frac{1}{ Z(\omega)}\right\}. \label{eq:chi_Z}
\eeq

\begin{figure}
	\centering
	\includegraphics[width=0.49\textwidth]{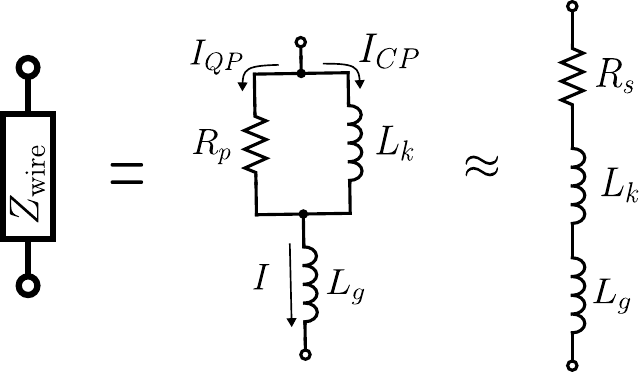}
	\caption{Circuit for a superconducting wire segment in the two-fluid model, where the nondissipative Cooper-pair current $I_{CP}$ is assumed to flow in parallel to the dissipative quasiparticle current $I_{QP}$. These are subjected to a kinetic inductor $L_{k}$ and a parallel resistor $R_{p}$, respectively. Both currents are subjected to the same geometric inductance $L_{g}$, as shown by $I=I_{CP}+I_{QP}$ in the circuit. An alternative approximate representation of the circuit assumes a quasiparticle resistor $R_s$ in series, satisfying $R_{s}\ll R_{p}$. A capacitor can be added to the circuit to describe a segment of a superconducting transmission line.} 
\label{fig:Wire}	
\end{figure}

Here we use the ``two-fluid model'' to describe the impedance of superconducting wires \cite{VanDuzer1999}. It assumes each wire segment is a kinetic inductor $L_{k}=\frac{\ell}{\omega \sigma_{2} A}$ \emph{in parallel} with a resistor $R_{p}=\frac{\ell}{\sigma_{1} A}$ as shown in Fig.~\ref{fig:Wire}. The former describes the flow of nondissipative Cooper-pair current ($I_{CP}$), while the latter describes dissipative quasiparticle current ($I_{QP}$). Here $\sigma(\omega) = \sigma_{1}+i\sigma_{2}$ is the complex conductivity of each wire segment,  and $\ell$ and $A$ are their length and cross-section area, respectively. 
A geometric inductance $L_{g}$ is added in series with this circuit in order to equally describe the external magnetic energy for both $I_{CP}$ and $I_{QP}$. The total impedance of the wire is then given by
\begin{equation}
    Z_{{\rm wire}}(\omega)=\frac{\ell}{\sigma(\omega)A}-i\omega L_{g}.
    \label{eq:impedance}
\end{equation}
Because in a superconductor $\sigma_{1}\ll \sigma_{2}$, this impedance is well approximated by
\begin{equation}
    Z_{{\rm wire}}(\omega)\approx \frac{\sigma_{1}\ell}{\sigma_{2}^{2}A}-i\omega\left(L_{k}+L_{g}\right), \label{eq:impedance_series}
\end{equation}
showing that the wire can also be represented by a resistor \emph{in series} with the inductances $L_k, L_g$, with resistance given by $R_{s}=\frac{\sigma_{1}\ell}{\sigma_{2}^{2}A}\ll R_{p}$, see Fig.~\ref{fig:Wire}. This is the circuit used in superconducting transmission lines \cite{VanDuzer1999}. 

A convenient expression for the total kinetic inductance can be obtained by noting that the penetration depth $\lambda$ is determined by equating measurements of $\sigma_{2}$ to the London expression $\sigma_{2}^{{\rm London}} = \frac{1}{\mu_0 \lambda^{2} \omega}$ where $\mu_0$ is the permeability of vacuum \cite{Hong2013}. This leads to
\beq \label{eq:Lk}
    L_k = \mu_0\lambda^{2}\frac{\ell}{A}.
\eeq

\subsection{Calculation of the conductivities}

The frequency-dependent conductivities $\sigma(\omega)=\sigma_1(\omega)+i\sigma_{2}(\omega)$ of a Bardeen-Cooper-Schrieffer (BCS) superconductor 
were calculated by Mattis and Bardeen in \cite{Mattis-Bardeen1958} for the case where the quasiparticles are at thermal equilibrium, with average occupation given by Fermi-Dirac functions $f(E)=1/(e^{E/k_BT}+1)$. 
Here $E\equiv E_k=\sqrt{\xi_{k}^{2}+\Delta^2}$ is the BCS quasiparticle energy, with $\xi_k=\epsilon_k-\epsilon_F$ the free eletron energy measured from the Fermi level $\epsilon_F$.

The Mattis-Bardeen theory can be generalized to the case away from thermal equilibrium, provided that quasiparticle occupation remains a function of QP energy $E$. We shall make this key \emph{QP energy distribution assumption} and
refer to the quasiparticle occupations as $n(E)$, a function that can differ from the equilibrium Fermi-Dirac function $f(E)$. The Mattis-Bardeen conductivities then become
\begin{subequations}
\beqa 
   \frac{\sigma_1}{\sigma_N} &=& \frac{2}{\hbar \omega} \int_{\Delta}^{\infty} dE \frac{E(E + \hbar \omega) + \Delta^2}{\sqrt{E^2 - \Delta^2} \sqrt{(E + \hbar \omega)^2 - \Delta^2}} \nonumber \\ 
   &\times& \left[n(E) - n(E + \hbar \omega) \right],\label{eq:sigma1}\\
  \frac{\sigma_2}{\sigma_N} &=& \frac{1}{\hbar \omega} \int_{\Delta - \hbar \omega}^{\Delta} dE \frac{E(E + \hbar \omega) + \Delta^2}{\sqrt{\Delta^2 - E^2} \sqrt{(E + \hbar \omega)^2 - \Delta^2}} \nonumber \\ 
  &&\times \left[ 1 - 2n(E + \hbar \omega) \right].\label{eq:sigma2}
\eeqa
\end{subequations}
These expressions are valid at subgap frequencies $\hbar\omega<2\Delta$, with $\sigma_N$ the non-SC (normal state) real part of the conductivity.


In order to connect to experiments, it is fruitful to express $\sigma_1(\omega)$ in terms of the number of QPs divided by the number of electrons bound as Cooper pairs \cite{Gao2008thesis,Martinis2009, Catelani2011},
\beq
x_{{\rm QP}}=\frac{N_{{\rm QP}}}{2\rho\Delta}=\frac{1}{\Delta}\int_{-\infty}^{\infty} d\xi \; n(\sqrt{\xi^2+\Delta^2}),
\label{eq:xQPexact}
\eeq
where $\rho$ is the electron energy density near $\epsilon_F$. 
As we shall argue below, the QP occupation appears to follow a ``quasithermal'' law $n(E)\approx n_0e^{-E/k_BT}$ in experiments. 
Here $n_0$ does not depend on QP energy $E$ or frequency $\omega$, but it may depend on other parameters such as temperature $T$ and gap $\Delta$. For the special case of thermal equilibrium we get $n_0=1$, as can be seen from $n(E)=f(E)=1/(e^{E/k_BT}+1)\approx e^{-E/k_BT}$. 
When the quasithermal law is followed and $k_BT\ll \Delta$, 
the QP density of states can be expanded around its singularity at $E=\Delta$:  $E/\sqrt{E^2-\Delta^2}\approx \sqrt{\Delta/[2(E-\Delta)]}$. Under this approximation we get
\beq
x_{{\rm QP}}\approx \sqrt{2}\int_{0}^{\infty}dx \frac{n\left((1+x)\Delta\right)}{\sqrt{x}}
=n_0\sqrt{\frac{2\pi k_{B}T}{\Delta}}e^{-\frac{\Delta}{k_BT}}.
\label{eq:xqpapprox}
\eeq


When both $\hbar\omega$ and $k_BT$ are much smaller than $\Delta$, the conductivity is also dominated by the singularity in the QP density of states; as a result, the same approximation as in Eq.~(\ref{eq:xqpapprox}) leads to the following analytic approximation for the real part of the conductivity:
\begin{eqnarray}
\frac{\sigma_1}{\sigma_N}&\approx&
x_{{\rm QP}}\left(\frac{2\Delta}{k_BT}\right)^{3/2}\frac{1}{\sqrt{\pi}}\left(\frac{k_BT}{\hbar\omega}\right)
\sinh{\left(\frac{\hbar\omega}{2k_BT}\right)}\nonumber\\
&&\times K_0\left(\frac{\hbar\omega}{2k_BT}\right),
\label{eq:sigma_1Approx}
\end{eqnarray}
where $K_0(y)$ is the modified Bessel function of the second kind. 

\begin{figure}
	\centering
	\includegraphics[width=0.49\textwidth]{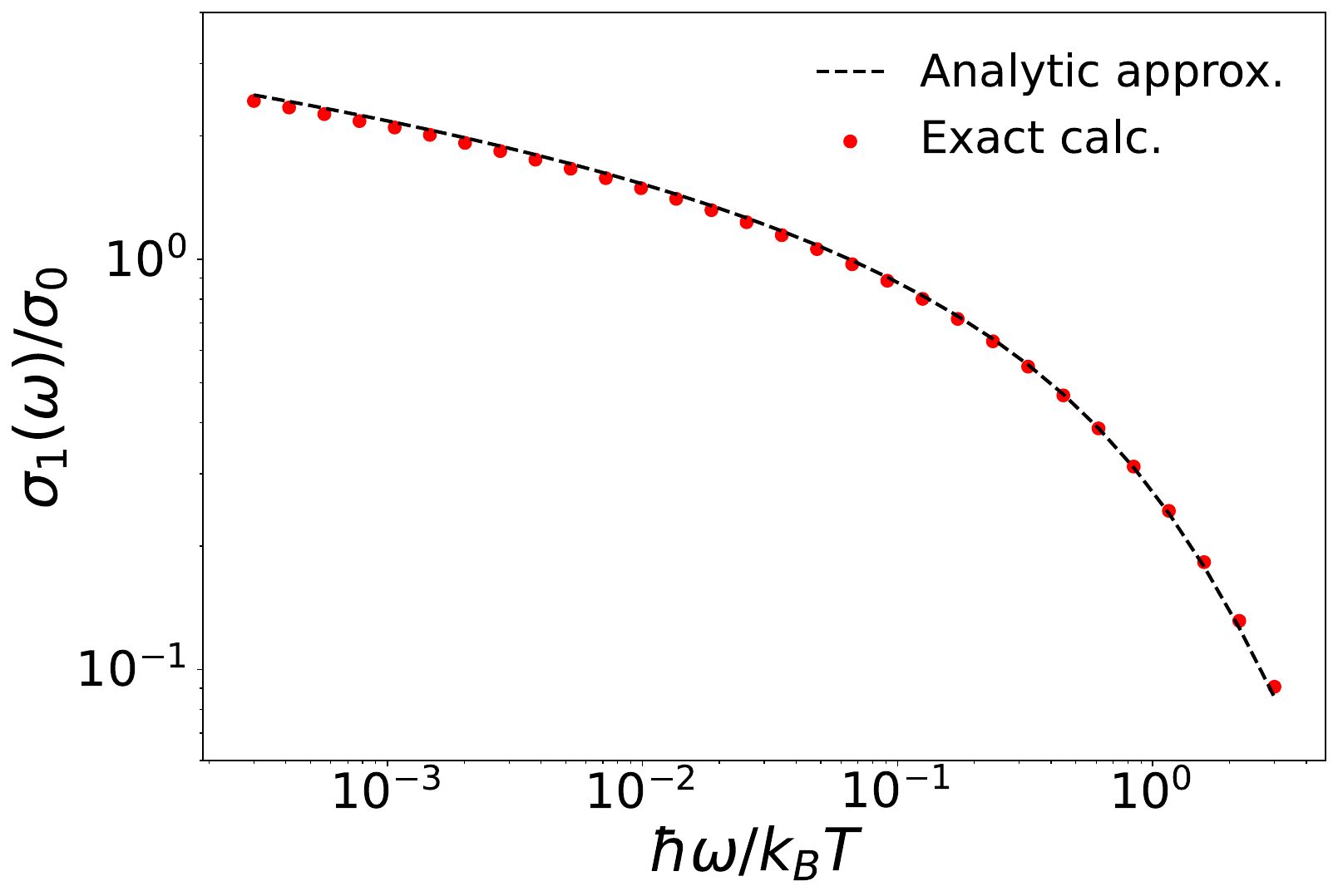}
	\caption{Numerical calculation of $\sigma_1(\omega)$ assuming the quasithermal law for the QP distribution observed in experiments, $n(E)\propto e^{-E/k_BT}$. The plot is normalized by $\sigma_0=\sigma_N x_{{\rm QP}}\left(2\Delta/k_BT\right)^{3/2}$. In the low frequency range $\hbar \omega \lesssim k_BT$, $\sigma_1$ decreases logarithmically with increasing $\omega$;  in the high frequency range it decreases as a power law. When $k_BT \lesssim 0.1 \Delta$ we find that the exact numerical result (red points) is well approximated by the analytical expression Eq.~(\ref{eq:sigma_1Approx}) (shown as a solid line for comparison).} 
\label{fig:sigma1}	
\end{figure}

Figure~\ref{fig:sigma1} compares Eq.~(\ref{eq:sigma_1Approx}) to exact numerical integration of both Eqs.~(\ref{eq:sigma1})~and~(\ref{eq:xQPexact}), assuming the quasithermal law and $k_BT/\Delta=0.1$. We find that Eq.~ (\ref{eq:sigma_1Approx}) approximates the exact result quite well provided that $\hbar\omega, k_BT \lesssim 0.1 \Delta$.

Thus, the behaviour of $\sigma_1(\omega)$ depends critically on the value of frequency relative to the thermal frequency $k_BT/\hbar$. 
In the low frequency regime of $\hbar\omega\ll k_BT$, $\sigma_1$ is logarithmic in frequency as
\beq \label{eq:sigma1LF}
\frac{\sigma_1}{\sigma_N}\approx \frac{1}{2\sqrt{\pi}}x_{{\rm QP}}\left(\frac{2\Delta}{k_BT}\right)^{3/2}\left[
\ln{\left(\frac{4k_BT}{\hbar\omega}\right)}-\gamma_E
\right],
\eeq
where $\gamma_E=0.5772\ldots$ is the Euler-Mascheroni constant. 
In the opposite high-frequency regime of $\hbar\omega\gg k_BT$, 
$\sigma_1$ is instead a power law, 
\beq \label{eq:sigma1HF}
   \frac{\sigma_1}{\sigma_N} \approx \frac{1}{2} x_{\rm QP} \left(\frac{2 \Delta}{\hbar \omega}\right)^{3/2}.
\eeq

The imaginary part of $\sigma$ has even simpler behaviour, because when $k_BT\ll \Delta$ we may assume $[1-2n(E+\hbar\omega)]\approx 1$ in Eq.~(\ref{eq:sigma2}). Thus when $\hbar \omega, k_BT \ll \Delta$, we can integrate Eq.~(\ref{eq:sigma2}) analytically to obtain
\beq \label{eq:sigma2_approx}
   \frac{\sigma_2}{\sigma_N} \approx \frac{\pi \Delta}{\hbar \omega}.
\eeq
Note how this agrees qualitatively with the phenomenological London theory. Therefore, it makes sense to equate $\sigma_{2}^{{\rm London}}$ to Eq.~(\ref{eq:sigma2_approx}) in order to obtain
\beq \label{eq:sigmaN}
    \sigma_N = \frac{\hbar}{\mu_0 \lambda^2 \pi \Delta}.
\eeq
This relation, valid for $\hbar \omega, k_BT \ll \Delta$, provides a practical method for computing $\sigma_N$.



We now combine these results with Section~\ref{subsec:wire} in order to derive useful expressions for characterizing superconducting circuits. First, assume a uniform wire such as a capacitively-coupled waveguide (CPW) or transmission line. Its resistance per unit length can be obtained from Eqs.~(\ref{eq:impedance_series}) and (\ref{eq:sigma2_approx}),
\begin{equation}
    \mathcal{R}_{{\rm wire}}=\frac{{\rm Re}\left\{Z_{{\rm wire}}(\omega)\right\}}{\ell}=\frac{\sigma_1}{\sigma_{2}^{2}A} = \frac{\sigma_1}{\sigma_N}\frac{\hbar\omega^{2}L_k}{\pi \Delta\ell}, 
    \label{eq:RCPW}
\end{equation}
leading to the admittance
\begin{equation}
    \frac{1}{Z_{{\rm wire}}(\omega)} \approx \frac{\sigma_1}{\sigma_N}\frac{\hbar}{\pi\Delta}\frac{L_k}{\left(L_k+L_g\right)^{2}} + \frac{i}{\omega \left(L_k+L_g\right)}.
    \label{eq:wireadmittance}
\end{equation}
Here $L_k$ and $L_g$ are the total kinetic and geometric inductances of the wire.


Equation~(\ref{eq:wireadmittance}) can be compared to the Josephson junction admittance $1/Z_J(\varphi)$ obtained by the quantum theory of Cooper pairs and QP tunneling across the junction \cite{Catelani2011},
\beq \label{eq:junction_admittance}
   \frac{1}{Z_J(\varphi)} = \frac{\sigma_1}{\sigma_N} \frac{\hbar}{\pi\Delta}\frac{\cos^{2}{\left(\frac{\varphi}{2}\right)}}{L_J} + \frac{i\left|\cos{\left(\varphi\right)}\right|}{\omega L_J},
\eeq
where $\varphi$ is the phase difference of the Cooper-pair wave function across the junction. 
The real part of the junction admittance is nonzero due to Ohmic loss of QPs tunneling across the junction; interestingly, this is equal to zero when $\varphi=\pi$. 

Compare this to the real part of our wire admittance Eq.~(\ref{eq:wireadmittance}), to see that energy loss away from junctions is described by $\varphi=0$ and the effective inductance $(L_k+L_g)^{2}/L_k$ instead of $L_J$.

\Addition{When using Eq.~(\ref{eq:junction_admittance}) one may need to account for quantum fluctuations of $\varphi$ by a suitable average over the qubit's wavefunction $\psi_j(\varphi)\equiv\langle \varphi |j\rangle$, or more generally its density matrix. But for low qubit energy states $|j\rangle=|0\rangle, |1\rangle$ it is often a good approximation to set $\varphi$ in Eq.~(\ref{eq:junction_admittance}) with the value that minimizes the qubit's effective potential; we assume this is a good approximation in our calculations below.}

%

\section{Energy Distribution for QPs in Quasiequilibrium}

A recent experiment presented evidence that SC QPs are at thermal equilibrium with the phonon bath, despite their out-of-equilibrium density. 
In \cite{Connolly2024} the following empirical expression for the distribution of QPs in SC circuits was proposed, 
\beq \label{eq:quasi-eq_dist}
   n(E) = x_{\rm QP} \sqrt{\frac{\Delta}{2 \pi k_BT}} e^{-\frac{E - \Delta}{k_BT}}.
\eeq
Note that this expression follows the quasithermal law mentioned above; in fact Eq.~(\ref{eq:quasi-eq_dist}) is a Maxwell-Boltzmann distribution with \emph{out-of-equilibrium chemical potential} $\mu=\Delta$. The fraction of QPs (normalized by the density of Cooper pairs) is modeled as
\beq \label{eq:xqp}
   x_{\rm QP} = x_{\rm QP}^{\rm res} + \sqrt{\frac{2 \pi k_BT}{\Delta}} e^{-\frac{\Delta}{k_BT}},
\eeq
so that the first contribution $x_{\rm QP}^{\rm res}$ is the fraction of QPs that are out of thermal equilibrium, with the second contribution describing QPs at thermal equilibrium. At high temperatures ($k_BT \geq 100$ mK in \cite{Connolly2024}), thermal QPs were found to dominate $x_{\rm QP}$, in that $x_{\rm QP} \approx \sqrt{2 \pi k_BT / \Delta} e^{- \Delta/k_BT}$ leading to $n(E)\approx f(E)$.
However, at low temperatures, resident QPs with temperature independent density $x_{\rm QP}^{\rm res}$ were found to dominate. 




\section{Quality factor of a CPW resonator}

We now quantify how the presence of resident QPs \emph{in quasiequilibrium} can limit the quality factor $Q$ of a CPW resonator. As shown in Appendix~\ref{sec:GDFT}, the fluctuation-dissipation theorem remains valid for QP energy distributions that satisfy the ``quasithermal law'' such as Eq.~(\ref{eq:quasi-eq_dist}) and thus Eq.~(\ref{SqFD}) can be applied. Our calculation below relies heavily on this assumption. 

As shown in \cite{Gao2008thesis, Goppl2008}, the CPW half-wavelength resonator can be modeled by an effective lumped element parallel $RLC$ circuit. At frequencies close to the lowest mode (with wavelength equal to $2\ell$) we get $R = \frac{2 \mathcal{L}}{\ell \mathcal{R} \mathcal{C}}$, $L = \frac{2 \ell \mathcal{L}}{\pi^2}$, and $C = \frac{\ell \mathcal{C}}{2}$ \cite{Gao2008thesis, Goppl2008}. Here $\mathcal{R},\mathcal{L},\mathcal{C}$ are the CPW's resistance, inductance, and capacitance per unit length, respectively. Use Eqs.~(\ref{Qtan}), (\ref{eq:chi_Z}) and ${\rm Re}\left\{\frac{1}{Z_{RLC}}\right\}=\frac{1}{R}$ to get
\begin{equation}
    \frac{1}{Q_{QP}}= \frac{1}{\Omega RC} = \frac{\mathcal{R}_{{\rm wire}}}{\Omega \mathcal{L}}.
\label{Qtan_CPW}
\end{equation}
%
%
Plugging Eq.~(\ref{eq:RCPW}) for $\omega=\Omega$, 
\begin{equation}
    \frac{1}{Q_{QP}}=\frac{\sigma_1(\Omega)}{\sigma_N}\frac{L_k}{L} \frac{\hbar \Omega}{\pi \Delta},
\label{Qtan_QP}
\end{equation}
where $L=\mathcal{L}\ell= L_k + L_g$ is the total inductance of the CPW. At high frequencies, 
\begin{equation}
    \frac{1}{Q_{QP}} \approx x_{{\rm QP}}\frac{L_k}{L} \left(\frac{2 \Delta}{\pi^2\hbar \Omega}\right)^{1/2}.
\label{Qtan_QP_hf}
\end{equation}

%

\begin{figure}
	\centering
	\includegraphics[width=0.45\textwidth]{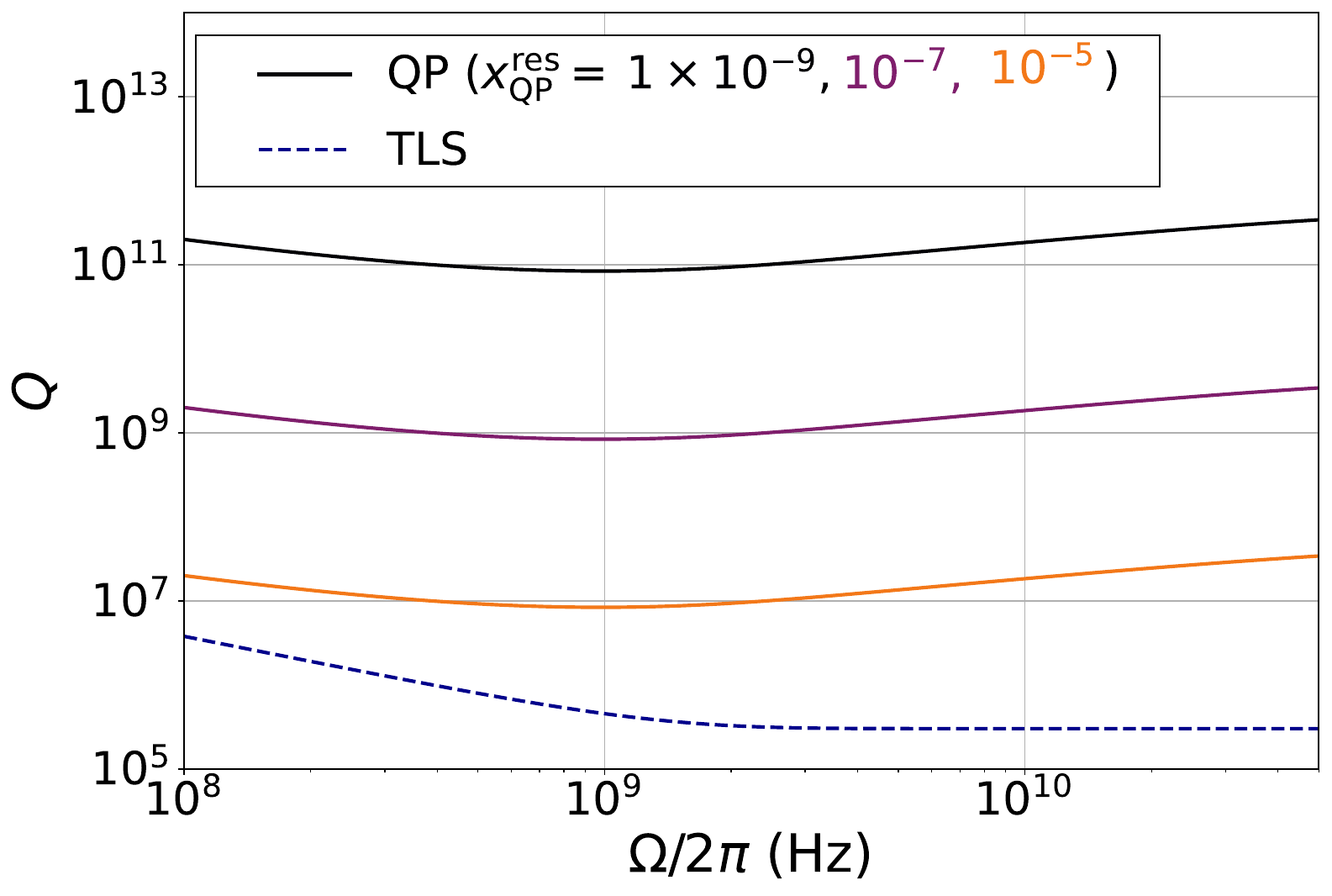}
\caption{Quality factor due to quasiequilibrium QPs (solid lines) in an aluminum CPW resonator for 3 values of $x_{\rm QP}^{\rm res}$ ranging from $10^{-9}$ (black) to $10^{-5}$ (orange). 
The length of the resonator is adjusted to match the resonance frequency $\Omega$ for the fundamental mode, $\ell=\frac{c \pi}{n \Omega}$.
For comparison, we include the $Q$ due to dielectric loss with typical loss tangents in bulk and surface (dashed line). Other parameters in Table \ref{table1}.\label{fig:Q_CPW}}
\end{figure}

Figure~\ref{fig:Q_CPW} shows calculations of $Q_{QP}$ as a function of CPW resonance frequency, $\Omega=\frac{1}{\sqrt{LC}}=\frac{1}{\sqrt{\mathcal{L}\mathcal{C}}}\frac{\pi}{\ell}$ (the lowest frequency mode). We used Eqs.~(\ref{eq:xqp}) and (\ref{Qtan_QP}) for $x_{\rm QP}^{\rm res}=10^{-9} - 10^{-5}$, the same values measured in aluminum qubits \cite{deVisser2011,Serniak2018, Connolly2024}. 

For each $\Omega$ we assume a CPW length equal to $\ell = \frac{c \pi}{n \Omega}$, where $c$ is the speed of light, and $n=\sqrt{11.7}$ is the substrate refractive index (silicon). 
We fixed the impedance $Z_0 = \sqrt{\mathcal{L}/\mathcal{C}}=50$~Ohms, so that 
$L=L_k+L_g = \mathcal{L}\ell=\pi Z_0/\Omega$, and used $L_k=\mu_0\lambda^{2}\ell/A$ (See other parameters in Table \ref{table1}). 


We find that a large density of resident QPs $x_{\rm QP}^{\rm res} \sim 10^{-5}$ limits $Q_{QP}$ to less than 
$10 ^{7}$ for aluminum CPWs with resonant frequencies in the GHz range. 

When subjected to different loss mechanisms, the quality factor of a resonator is given by $Q^{-1}= \sum_i Q_i^{-1} + Q_{QP}^{-1}$, where the contribution due to Ohmic loss from QPs is $Q_{QP}^{-1}$ and the $Q_i^{-1}$s are contributions from other loss mechanisms. An important mechanism is the contribution from dielectric loss due to TLSs, denoted $Q_{TLS}$ in Fig.~\ref{fig:Q_CPW}. We evaluated $Q_{TLS}$ by including a surface loss tangent in Eq.~(\ref{avgtandeltaTLS}) in order to account for (1) TLSs in the dielectric on the surface of the SC wire forming the center-strip of the CPW (metal-air and metal-substrate interfaces described in \cite{Wang2015, Murray2020}), plus (2) the dielectric at the surface of the substrate in the gap between the center-strip and the ground plane (substrate-air interface). In addition, we also included a bulk loss tangent due to TLSs at the substrate. This led to a total value of $Q_{TLS}$= 3 $\times 10^{5}/\tanh{\left(\hbar\Omega/2k_BT\right)}$ which is valid in the low power regime (when TLS absorption is \emph{not saturated} \cite{Gorgichuk2023}). 

For current materials Fig.~\ref{fig:Q_CPW} shows that $Q_{TLS}<Q_{QP}$, so that the $Q$ of aluminum CPWs is dominated by TLS loss. However, a modest amount of TLS mitigation may make the QP mechanism  relevant. In particular, we expect $Q\approx Q_{QP}$ in measurements of $Q$ at high power when the TLS absorption is saturated.

 

\begin{table*}[t]
\centering
\caption{\textbf{Device parameters used in figures.} \\ $T$=30 mK, $T_c=1.2$~K, $\Delta/(2\pi\hbar)=44$~GHz, $\lambda=50$~nm (aluminum), $\tan (\delta^S_{TLS, 0})$=$1\times 10^{-3}$, and $\tan (\delta^B_{TLS, 0})$=$1\times 10^{-6}$ in all cases.}
\label{tab:example_table}
\begin{tabular}{lcccc}
\toprule
 
 & Fig. \ref{fig:Q_CPW} & Fig. \ref{fig:T1_transmon} & Fig. \ref{fig:Sphi} & Fig. \ref{fig:T2FID}\\
\midrule
CPW center-strip ($\mu m$)  & 10 &  & 10 & \\
CPW gap ($\mu m$)  & 6 &  & 6 & \\
$\ell$ ($\mu$m) & $\frac{c \pi}{n \Omega}$ & 15 (lead), 500 (pad) & 1200 & \\
A ($\mu$m$^2$) & 1 & \Addition{0.1 (lead), 25 (pad)} & 1 & 0.01\\
$p_S$ ($\times 10^{-4}$) \cite{Wang2015, Murray2020}  & 23 & 2.4 & &\\
$p_B$ \cite{Wang2015}  & 0.9 & 0.9 &  &\\
$\hbar \Omega$ & $\hbar \pi/\sqrt{LC}$ &  $\sqrt{8E_J E_C}$ & & $\sqrt{8\bar{E}_J(\Phi) E_C}$\\
$Z_0=\sqrt{\mathcal{L}/\mathcal{C}}$ (Ohms) & 50 & & &\\
$E_J/E_C$ &  & 70 & &70\\
$C$ (pF) & $\pi/(Z_0\Omega)$ & $\sqrt{2e^4E_J/E_C}/(\hbar\Omega)$ & 0.1 &\\
$L$ & $L_k + L_g$& $L_J + 2L_k + 2L_g$ & $\left[ 1/L_J + 1/(L_k+L_g)\right]^{-1}$  & $L = \frac{L_{J}}{2\cos{\left(\pi\frac{\Phi}{\Phi_0} \right)}} + L_k + L_g$\\
$L_k$ (nH) & $3.15\times 10^{-6} \frac{\ell}{\mu{\rm m}}$ & 4.7 $\times 10^{-4}$ & 0.003 & 1.6$\times 10^{-4}$\\
$L_g$ (nH) & $\pi Z_0/\Omega -  L_k$& 2 $\times 10^{-2}$ & 0.6  & \\
$L_J$ (nH) & & $\left(\frac{\Phi_0}{2\pi}\right)^2 \sqrt{8E_c/E_J}/(\hbar\Omega)$ & 0.24  & 10\\

\bottomrule
\end{tabular} \label{table1}
\end{table*}

\section{Energy relaxation of a transmon qubit}
\begin{figure}[htbp]
  \centering
  \subfloat[\label{fig:circuit1}]{
    \includegraphics[width=0.18\textwidth]{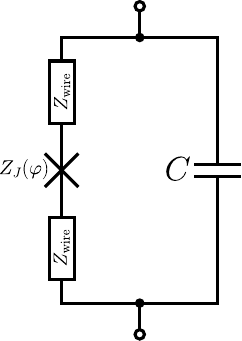}
  }
  \hfill
  \subfloat[\label{fig:circuit2}]{
    \includegraphics[width=0.24\textwidth]{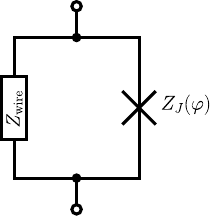}
  }

  \par\vspace{1em} 

  \subfloat[\label{fig:circuit3}]{
    \includegraphics[width=0.44\textwidth]{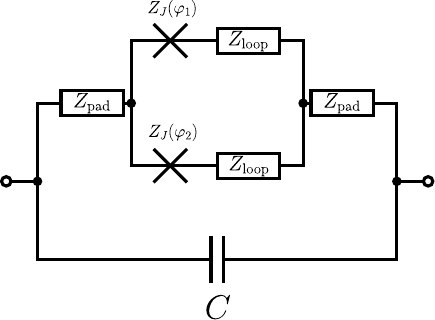}
  }
  \caption{Circuit representation of (a) transmon, (b) flux qubit, and (c) split-junction transmon. Each wire includes the dissipative resistance due to Ohmic loss in addition to the inductive response as explained in Fig. \ref{fig:Wire}. In the transmon, each $Z_{\rm wire}$ includes a small lead connected to the junction and the large pad electrode forming the capacitor. In the flux qubit $Z_{\rm wire}$ represents the wire loop. For the split-junction transmon, each $Z_{\rm loop}$ represents one half of the wire loop with the pads considered separately in $Z_{\rm pad}$.}
  \label{fig:Circuits}
\end{figure}

We now calculate the relaxation rate $1/T_1$ due to quasiequilibrium QPs in a transmon. We account for two dinstinct mechanisms of QP loss: QP tunneling across the Josephson junction and QP Ohmic loss within the wires. 

The transmon is formed by two electrodes connected on each side of a Josephson junction, i.e. electrode-junction-electrode in series \cite{Wang2015,Connolly2024, McEwen2024}. Each electrode consists of a small lead connected to a large pad, see e.g. Fig. 1 of \cite{Wang2015}. Typically, the  lead has the \Addition{largest} $\ell/A$ of all other wire segments forming the electrode, and as a result the lead dominates $L_k$. 
In contrast, the pad \Addition{has much larger footprint} so it dominates $L_g$. In this section we assume a transmon with the same geometry of Fig. 1(a) of \cite{Wang2015}, the circuit considered is shown in Fig. \ref{fig:circuit1}.

The impedance of the nongap-engineered (NGE) transmon is then $Z_{\rm transmon}^{{\rm NGE}}=Z_J(0)+2 Z_{{\rm wire}}$, where $Z_J(0)$ is the impedance of the junction with $\varphi=0$ \Addition{(the minimum of the effective potential for the transmon \cite{Koch2007})} and $Z_{{\rm wire}}$ is the impedance of each electrode. Using Eqs.~(\ref{eq:impedance_series}) and (\ref{eq:junction_admittance}) with $\varphi=0$, the admittance of an NGE transmon is given by:
\begin{eqnarray}
    \frac{1}{Z_{\rm transmon}^{{\rm NGE}}}&\approx& \frac{\sigma_1}{\sigma_N}\frac{\hbar}{\pi \Delta}\frac{\left(L_J+2L_k\right)}{\left[L_J+2\left(L_k +L_g\right)\right]^{2}}\nonumber\\
    &&+\frac{i}{\omega\left[L_J+2\left(L_k +L_g\right)\right]}. \label{eq:ZtransmonNGE}
\end{eqnarray}
%
%
%
Note that we have not included the capacitance's impedance. A lossless capacitor in parallel will not have an impact on the real part of the total admittance. This is the case for the remaining qubits considered in this work; thus, we will not account for this contribution. Instead, we separately account for the capacitive loss due to TLSs and compare with QP loss as in the CPW case. From Eqs. (\ref{T1chi}) and (\ref{eq:chi_Z}),
\begin{eqnarray}
    \frac{1}{T_{1}^{{\rm NGE}}} &=& \frac{\sigma_1(\Omega)}{\sigma_N} \frac{\hbar}{C\pi \Delta} 
    \frac{\left(L_J + 2 L_k\right)}{\left[L_J+2\left(L_k + L_g\right)\right]^{2}} \nonumber \\
    &\times& \coth{\left(\frac{\hbar\Omega}{2k_BT}\right)}. 
\label{T1_QP_transmon}
\end{eqnarray}
For $L_J \gg 2(L_k + L_g)$, we recover the relaxation rate due to QPs tunneling across a Josephson junction \emph{without gap asymmetry} \cite{Martinis2009, Catelani2011}:
\begin{equation}
    \frac{1}{T_{1}^{{\rm NGE}}} \approx x_{\rm QP}  \left(\frac{2 \Delta \Omega}{\hbar \pi^2}\right)^{1/2},
\label{T1_QP_junction}
\end{equation}
valid at high frequencies with $\hbar \Omega \approx  \hbar/\sqrt{L_JC}$.

For gap engineered (GE) transmon qubits, where the gap asymmetry is larger than the qubit's frequency, the real part of the junction admittance is exponentially suppressed \cite{Marchegiani2022, Connolly2024}. In this case, the total admittance is dominated by Ohmic loss in the lead. The relaxation rate due to Ohmic loss from QPs in GE transmons is then given by
\begin{eqnarray}
    \frac{1}{T_{1}^{{\rm GE}}} &=& \frac{\sigma_1(\Omega)}{\sigma_N} \frac{\hbar}{C\pi\Delta}  \frac{\left(2 L_k\right)}{\left[L_J+2\left(L_k + L_g\right)\right]^{2}}\nonumber \\
    &\times& \coth{\left(\frac{\hbar\Omega}{2k_BT}\right)}.  
\label{T1_QP_GEtransmon}
\end{eqnarray}
For $L_J \gg 2(L_k + L_g)$ and high frequencies, we get
\begin{equation}
    \frac{1}{T_{1}^{GE}} \approx \frac{2L_k}{L_J} x_{\rm QP}  \left(\frac{2 \Delta \Omega}{\hbar \pi^2}\right)^{1/2}.
\label{T1_QP_hf}
\end{equation}

\begin{figure}
	\centering
	\includegraphics[width=0.45\textwidth]{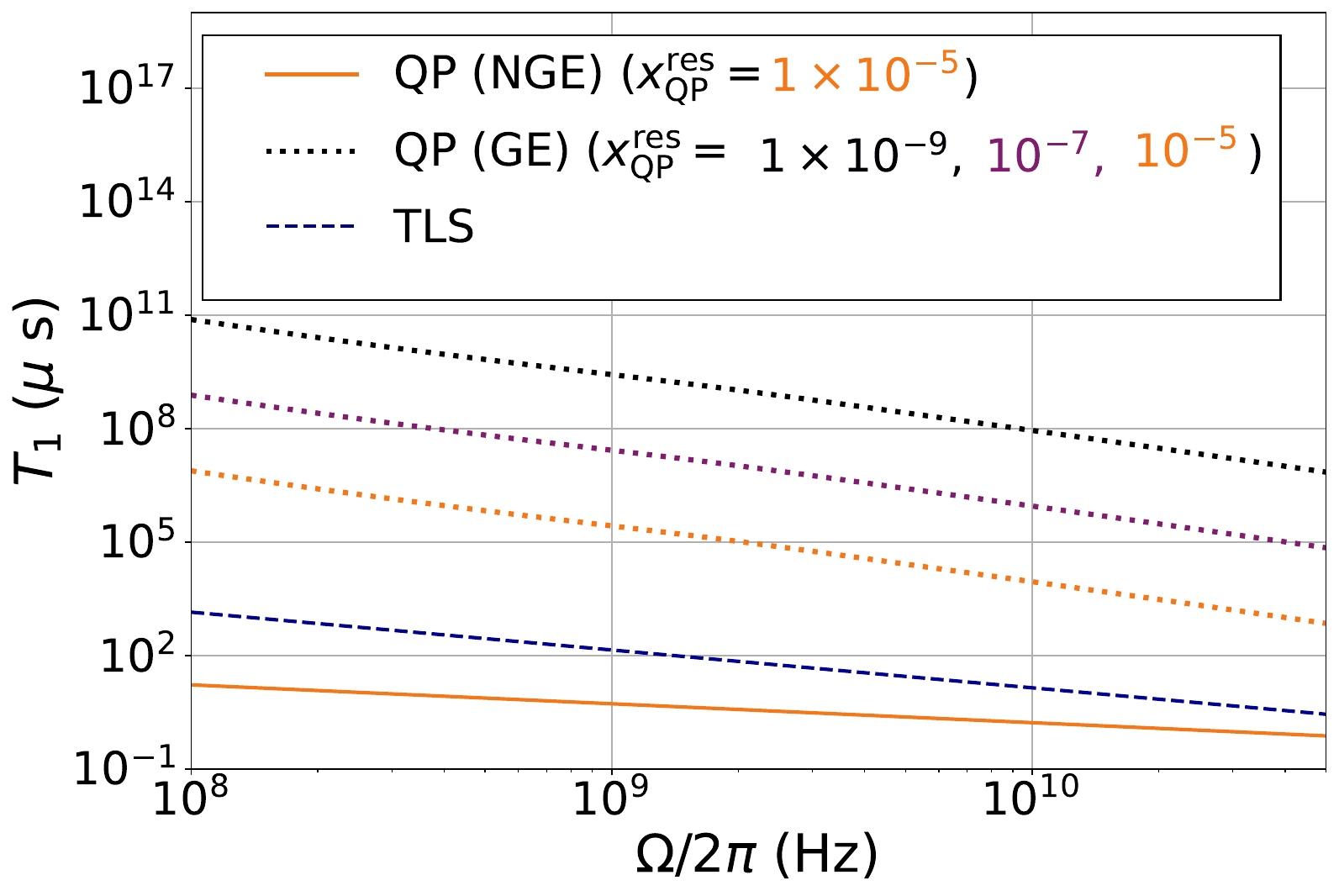}
\caption{Energy relaxation time $T_1$ due to quasiequilibrium QPs in a single-junction aluminum transmon. Solid line: $T_1$ for a non-gap engineered (NGE) transmon for $x_{\rm QP}^{\rm res}=10^{-5}$, which is dominated by QP tunneling across the junction. Dotted lines: $T_1$ for gap-engineered (GE) transmon for 3 values of $x_{\rm QP}^{\rm res}$. For comparison, the dashed-blue-line shows $T_1$ due to dielectric loss assuming typical loss tangents in bulk and surface.}
\label{fig:T1_transmon}
\end{figure}

Figure~\ref{fig:T1_transmon} shows the $T_1$ for NGE (\ref{T1_QP_transmon}) and GE (\ref{T1_QP_GEtransmon}) aluminum transmons as a function of qubit frequency $\Omega$ for a few different $x_{\rm QP}^{\rm res}$.  From Eq.~(\ref{eq:Lk}) we estimate $L_k\approx \mu_0 (50~{\rm nm})^{2}\times 15\mu{\rm m}/(0.1\mu {\rm m}^{2})=0.47$~pH; and $L_g\sim \mu_0\ell_{{\rm pad}}=\mu_0 \times 500 \mu{\rm m}=0.6$~nH. Note how the electrode inductances are much smaller than the junction inductance $L_J= 10-100$~nH. We fix the ratio $E_J/E_c = 70$, and vary $C$ and $L_J$ according to Table~\ref{table1} to match $\Omega$. Other parameters are shown in Table \ref{table1}. For comparison, we also plot the $T_1$ from dielectric loss due to TLSs. 

At the largest $x_{\rm QP}^{\rm res}$, the $T_1$ of NGE transmons is dominated by resident QPs tunneling across the junction, limiting $T_1$ to tens of $\mu$s as suggested by experiments \cite{McEwen2024, Connolly2024}.


In GE transmons, this QP-tunneling mechanism is shut-off and $1/T_1$ due to QPs is reduced by several orders of magnitude. For example, the reduction factor is $2L_k/L_J\sim 10^{-4}$ for aluminum transmons with $\frac{\Omega}{2\pi}=5$~GHz. Therefore, in GE transmons TLSs dominate the energy loss.



\section{Flux noise from quasiparticles in quasiequilibrium}

As explained at the end of Section~\ref{sec:relaxation_noise}, there is a flux noise spectral density associated with charge noise, as in $\tilde{S}_{\Phi}(\omega) =  (L \omega)^2\tilde{S}_Q(\omega)$ where $L$ is the total inductance of the wire. In this section, we calculate the flux noise contribution from QPs in two scenarios of interest: (1) the flux noise in a flux qubit, which we compare to the flux noise extracted from $T_1$ measurements in such qubits; and (2) the flux noise contribution to decoherence time $T_2^{*}$ in flux-tunable split-junction transmons.

\subsection{Flux noise spectral density in flux qubits}

Several experiments extract the environmental flux noise spectral density from relaxation time measurements in flux-sensitive qubits using Eq. (\ref{T1genPhi}) \cite{Quintana2017, Luthi2018, Stavengathesis}. Here we calculate and compare the flux noise contribution from QPs in a flux qubit to experimental results. This subsection considers a Superconducting Quantum Interference Device (SQUID) acting as a flux qubit formed by a SC wire loop with a single Josephson junction, as shown in Fig.~\ref{fig:circuit2}. In this case, the Hamiltonian of the flux qubit is given by
\begin{equation}
    {\cal H}=\frac{E_C}{e^2}Q^2-E_J \cos \left(2\pi \frac{\Phi}{\Phi_0}\right) + \frac{E_L}{2}\left( \frac{2\pi}{\Phi_0} \right)^2 \left(\Phi - \Phi_X\right)^2, \label{eq:Hamiltonian_FQ}
\end{equation}
with $E_L = \left(\frac{\Phi_0}{2\pi}\right)^2 \frac{1}{L_g+L_k}$ and $L_k, L_g$  the loop's kinetic and geometric inductances, respectively. The qubit's potential can be controlled by the external flux 
$\Phi_X$. 
The flux qubit is realized when $\beta=E_J/E_L>1$. In this regime the qubit is described by a double well potential as a function of flux, which allows for the encoding of quantum information in superpositions of persistent current states flowing clock and anti-clockwise in the SQUID loop \cite{Mooij1999}. 

%
%

The circuit for the flux qubit is shown in Fig.~\ref{fig:circuit2}; note how the wire formed by the SC loop is in parallel to the junction. Consider an NGE junction and use Eqs. (\ref{eq:wireadmittance}) and (\ref{eq:junction_admittance}) to obtain the real part of the flux qubit total admittance,
\beq \label{eq:fluxqubit_admittance}
   {\rm Re} \left\{ \frac{1}{Z_{\rm FQ}\left(\varphi\right)} \right\} = \frac{\sigma_1}{\sigma_N} \frac{\hbar}{ \pi\Delta } \left[ \frac{L_k}{(L_k+L_g)^2} + \frac{\cos^{2}{\left(\frac{\varphi}{2}\right)}}{L_J}\right],
\eeq
with phase $\varphi = 2\pi \frac{\Phi}{\Phi_0}$. 
%
%
%
Using Eqs.~(\ref{SqFD}), (\ref{eq:SPhigen}), (\ref{eq:chi_Z}), and (\ref{eq:fluxqubit_admittance}) we get the associated flux noise due to resident QPs,
    \begin{eqnarray}
    \tilde{S}_{\Phi}^{\rm FQ}(\omega) &=&  \frac{\sigma_1(\omega)}{\sigma_N}\frac{2 \hbar^2 \omega L^2}{\pi\Delta} \left[ \frac{L_k}{(L_k+L_g)^2} + \frac{\cos^{2}{\left(\frac{\varphi}{2}\right)}}{L_J} \right] \nonumber \\ &\times& \left[ n_B(\omega) + 1 \right], \nonumber \\ \label{eq:Sphi}
\end{eqnarray}
with total inductance $L = \left[ 1/L_J + 1/(L_k+L_g)\right]^{-1}$. \Addition{Equation~(\ref{eq:Sphi}) should be averaged over the qubit's wavefunction $\psi_j(\varphi)$; in practice, $\varphi$ can be approximated by the minimum value of the flux qubit double well potential, which is $\varphi=\pi\pm \sqrt{6(\beta-1)}$, where $\beta=E_J/E_L$}. 
%
For the typical flux qubit $L_g\gg L_J,L_k$ making $L\approx L_J$ and flux noise linearly proportional to $L_J$.

Similarly, for a flux qubit with a GE junction, we get
\begin{eqnarray}
    \tilde{S}_{\Phi}^{\rm GE \ FQ}(\omega) &=&  \frac{\sigma_1(\omega)}{\sigma_N}\frac{2 \hbar^2 \omega L^2}{\pi\Delta} \left[ \frac{L_k}{(L_k+L_g)^2}\right] \left[ n_B(\omega) + 1 \right]. \nonumber \\ \label{eq:SphiGE}
\end{eqnarray}%
At low frequencies ($\hbar\omega \ll k_BT$), flux noise scales logarithmically with frequency and temperature, $\tilde{S}_{\Phi}(\omega) \propto \ln{\left(4k_BT/\hbar\omega\right)}/T^{1/2}$. In moderately sized frequency bands this will appear as ``nearly white'' flux noise. In contrast, for $\hbar\omega \gg k_BT$, $\tilde{S}_{\Phi}(\omega) \propto 1/\sqrt{\omega}$.

Luthi et al. \cite{Luthi2018} observed a white flux noise background of magnitude $3.6 \times 10^{-15} \ \Phi_0^2/ {\rm Hz}$ of unknown origin in a NbTiN SQUID. Another experiment performed in a different device of the same material measured a similar white flux noise background of magnitude $2 \times 10^{-16} \ \Phi_0^2/ {\rm Hz}$ \cite{Stavengathesis}. Can our proposed QP mechanism explain the origin of this white flux noise background? To try to answer this question, we use our 
Eq.~(\ref{eq:Sphi}) to estimate the required $x_{\rm QP}^{\rm res}$ that yields the noise level observed in these experiments. 
We approximate $L \approx L_J$, $\cos^{2}{\left(\frac{\varphi}{2}\right)} \approx 1$ for $\beta\approx 2$ and assume an effective $L_J \approx $ 5 nH, other parameters as in \cite{NoteNbTiN}. We find that $x_{\rm QP}^{\rm res}\approx 3 \times 10^{-4}$ in Eq.~(\ref{eq:Sphi}) would explain the white flux noise background obtained in both experiments \cite{Luthi2018, Stavengathesis}.

To our knowledge, $x_{\rm QP}^{\rm res}$ has not yet been measured in NbTiN devices. Instead, an experiment performed in a NbTi resonator measured $x_{\rm QP}^{\rm res} \approx 7 \times 10^{-6}$ \cite{Barends2011}. 
It's important to note that the QP density depends on the material but also on the specific sources driving the electron gas out of equilibrium. Note that a larger $L_J$ considered in a calculation would result in a smaller $x_{\rm QP}^{\rm res}$. A measurement of $x_{\rm QP}^{\rm res}$ in these devices is needed before reaching a conclusion.

\begin{figure}
	\centering
	\includegraphics[width=0.45\textwidth]{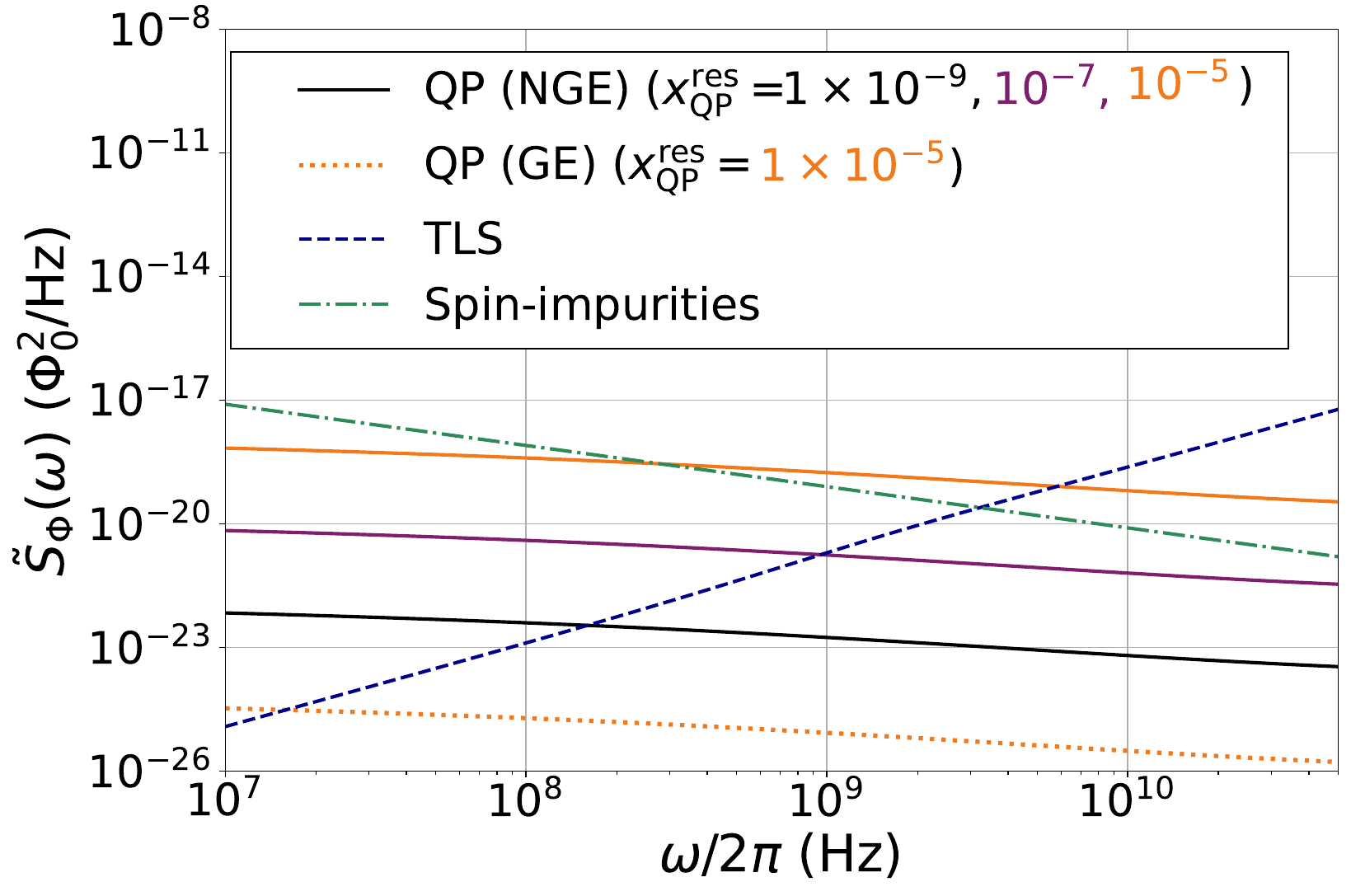}
	\caption{Flux noise due to quasiequilibrium QPs in a flux qubit with aluminum CPW, using parameters similar to experiment \cite{Quintana2017} (CPW length $\ell= 1.2$~mm and other parameters in Table~\ref{tab:example_table}). Solid lines: NGE flux qubit. Dotted line: GE flux qubit. Dashed line: Flux noise due to dielectric loss. Dash-dotted line: Flux noise due to spin impurities.} 
\label{fig:Sphi}	
\end{figure}

Figure~\ref{fig:Sphi} shows the flux noise contribution due to resident quasithermal QPs in an aluminum CPW of length $\ell=$ 1.2 mm acting as an inductor in an NGE flux qubit with $\beta =$ 2.5, as in Ref. \cite{Quintana2017}, with normalized QP density of $x_{\rm QP}^{\rm res} = 1\times10^{-9} - 1\times10^{-5}$. We assume $\cos^{2}{\left(\frac{\varphi}{2}\right)} = 1$, additional parameters in Table~\ref{table1}.

We compare this to a GE flux qubit with $x_{\rm QP}^{\rm res}=1\times10^{-5}$ and the same parameters. The flux noise amplitude is considerably smaller in flux qubits with GE junctions where QP tunneling is suppressed. Note that since $L_k + L_g \gg L_J$ in a GE flux qubit, flux noise can be significantly reduced by a factor of $L_JL_k/(L_k+L_g)^2$ compared to the NGE case as implied by Eq. (\ref{eq:SphiGE}).

Our predicted amplitudes of the ``nearly white'' flux noise background due to large densities of QPs are of the same order of magnitude as the amplitude measured in flux noise experiments by Quintana {\it et al.} \cite{Quintana2017}, in the $10-1000$~MHz band with aluminum devices of same dimension as in Fig.~\ref{fig:Sphi}. Quintana {\it et al.} \cite{Quintana2017} measured a flux noise frequency dependence that changed from $1/\omega$ at low frequency to $\omega^{m}$ with exponent $m=1-3$ at large frequencies. This frequency dependence can not be explained by the present QP mechanism. The low frequency $1/\omega$ contribution can be explained by spin impurities \cite{Nava2022}, while the high frequency ``superOhmic'' contribution remains unexplained. 

For comparison, Fig.~\ref{fig:Sphi} also shows the flux noise contribution due to spin-impurities $\tilde{S}^{\rm spins}_{\Phi}(\omega) = 16\pi \times 10^{-11} \Phi_{0}^{2}/\omega$ \cite{Quintana2017}. We also show the contribution due to TLSs in the dielectric materials of the CPW resonator,
\begin{eqnarray}
    \tilde{S}^{\rm TLS}_{\Phi}(\omega) &=&  (L \omega)^2\tilde{S}^{\rm TLS}_Q(\omega)\nonumber\\
    &=& 2 \hbar L^2 \omega^2 C \langle\tan{(\delta_{TLS})}\rangle \left[ n_B(\omega) + 1 \right].
    \label{SPhiTLS}
\end{eqnarray}
Here we point out that the flux noise associated to TLS charge noise may provide an explanation for the superOhmic noise measured in \cite{Quintana2017}. As we see in Eq.~(\ref{SPhiTLS}), $\tilde{S}^{\rm TLS}_{\Phi}(\omega)$ is proportional to $\omega^2$ in the low $T$ regime. A TLS loss tangent of $\langle\tan{(\delta_{TLS})}\rangle \approx 2 \times 10^{-4}$ as in Fig. \ref{fig:Sphi} would match the superOhmic flux noise measured by Quintana {\it et al.} with superOhmic exponent $m=2$.

\subsection{$T_2^*$ due to off-resonant flux noise in a split-junction transmon}

When the qubit frequency $\Omega$ is sensitive to flux fluctuations, 
off-resonant flux noise at $\omega\neq \Omega$ can significantly impact the decoherence times measured in Ramsey ($T_2^*$) and Hahn-echo ($T_2$) experiments. The average value of the qubit coherence operator $\sigma_+ \equiv \sigma_x + i\sigma_y$ evolves in time as \cite{deSousa2009}
\begin{eqnarray}  \label{eq:decayT2}
    |\langle \sigma_+(t) \rangle| &=& \exp{\left[- \int_{-\infty}^{\infty} d\omega \left(\frac{\partial\Omega}{\partial\Phi}\right)^2 \tilde{S}_{\Phi}(\omega) \mathcal{F}(\omega,t) \right]} \nonumber \\ &\equiv& \exp{\left[-\alpha(t)\right]},
\end{eqnarray}
where 
$\mathcal{F}(\omega,t)$ is a filter function that depends on the pulse sequence used in the experiment. A similar expression can be obtained in terms of charge noise.

In this subsection we consider a ``split-transmon qubit'' (ST), which has its JJ replaced by a SC loop with two JJs in parallel, see Fig.~\ref{fig:circuit3}. We assume the ST has two identical junctions, each with Josephson energy $E_J$, leading to the Hamiltonian
\begin{eqnarray}
        {\cal H} &=& \frac{E_C}{e^2}Q^2-E_J \cos \left(\varphi_1\right) -E_J \cos \left(\varphi_2\right)\nonumber\\ 
        &&+ \frac{E_L}{2} \left(\frac{2\pi}{\Phi_0}\right)^{2} \left(\Phi-\Phi_X\right)^{2},         
        \label{eq:Hamiltonian_ST}
\end{eqnarray}
where $\varphi_1$ and $\varphi_2$ are the phase differences across each junction and $\Phi, \Phi_X$ are the total flux and external flux coupled to the loop, respectively. 
These are related by $\Phi=\Phi_X+\frac{LI_c}{2}\left(\sin{\varphi_2}-\sin{\varphi_1}\right)$ and by the flux quantization relation $\Delta\varphi=\varphi_1-\varphi_2=2\pi\frac{\Phi}{\Phi_0}$. Using the latter and assuming 
the loop is small enough such that its associated inductive energy satisfies $E_L\gg E_J/4$ we get 
\begin{equation}
        {\cal H}\approx  \frac{E_C}{e^2}Q^2 -2 E_J\cos{\left(\pi\frac{\Phi}{\Phi_0}\right)}
    \cos(\varphi), \label{eq:Hamiltonian_STsimple}
\end{equation}
where $\varphi=\frac{1}{2}\left(\varphi_1+\varphi_2\right)$. In the transmon regime ($E_J\gg E_c$) this Hamiltonian is minimized for wave functions peaked at 
$\varphi_1 = \varphi+\frac{1}{2}\Delta\varphi=\pi\frac{\Phi}{\Phi_0}$, $\varphi_2 = \varphi-\frac{1}{2}\Delta\varphi=-\pi\frac{\Phi}{\Phi_0}$ ($\varphi_1 = \pi +\pi\frac{\Phi}{\Phi_0}$, $\varphi_2 = \pi -\pi\frac{\Phi}{\Phi_0}$) when $\cos \left( \pi\frac{\Phi}{\Phi_0} \right) >0$ ($\cos \left(  \pi\frac{\Phi}{\Phi_0} \right) <0$). 
The value of $\Phi$ that minimizes the energy is a nonlinear function of $\Phi_X$ that has to be calculated numerically; but since $\Phi$ is controllable by the external flux $\Phi_X$ we will assume $\Phi$ is the variable that provides control of the ST qubit.  

The ST behaves as a single-junction transmon with effective Josephson energy $\bar{E}_J(\Phi)$ controllable by $\Phi$; its resonance frequency is ``flux tunable''
\begin{equation}
    \hbar\Omega(\Phi)=\sqrt{8\bar{E}_J(\Phi)E_c}=
\hbar\sqrt{\frac{2\left|\cos{\left(\pi\frac{\Phi}{\Phi_0}\right)}\right|}{L_JC}},
\end{equation}
and is subject to decoherence due to off-resonant flux noise.

%
%

%
%


We now calculate the impedance of the ST circuit shown in Fig.~\ref{fig:circuit3}. 
We assume that the SC loop is formed by two segments of wire + junction in parallel, each segment with impedance $Z_{\rm loop} + Z_J(\varphi_1)$ and $Z_{\rm loop} + Z_J(\varphi_2)$. 
\Addition{Using the values of $\varphi_1,\varphi_2$ that minimize the split-transmon effective potential in Eq.~(\ref{eq:Hamiltonian_STsimple})} we get 
%
%
$Z_J(\varphi_1) = Z_J(\varphi_2) = Z_J(\phi)$, where $\phi=\pi\frac{\Phi}{\Phi_0}$ ($\phi=\pi+\pi\frac{\Phi}{\Phi_0}$) for $\cos \left(  \pi\frac{\Phi}{\Phi_0} \right) >0$ ($\cos \left(  \pi\frac{\Phi}{\Phi_0} \right) <0$). 
The total impedance in an ST with NGE junctions is thus $Z_{\rm ST}^{NGE} = 2Z_{\rm pad} + Z_{\rm loop}/2 + Z_J(\phi)/2$, where $Z_{{\rm loop}}$ is the impedance of half of the wire forming the loop and 
$Z_{\rm pad}$ is the impedance of each pad electrode. We can approximate the junction's impedance from Eq. (\ref{eq:junction_admittance}),
\beq \label{eq:junction_impedanceapprox}
   Z_J(\phi) \approx \frac{\sigma_1}{\sigma_N} \frac{\hbar}{\pi\Delta}\frac{\omega^2 L_J \cos^{2}{\left(\frac{\phi}{2}\right)}}{\cos^{2}{\left(\phi\right)}} - \frac{i \omega L_J}{\left|\cos{\left(
   \phi\right)}\right|}.
\eeq
This approximation is valid at subgap frequencies and $|\cos{\left(\phi\right)}|/ \cos^{2}{\left(
\phi/2\right)} \gg \frac{\sigma_1}{\sigma_N} \frac{\hbar \omega}{\pi\Delta}$. For the frequencies and QP densities of interest, it is safe to use this approximation away from $\phi = \pi/2, 3\pi/2$. Using this approximation and Eq.~(\ref{eq:impedance_series}) we calculate the ST total admittance with NGE junctions,
\begin{eqnarray}
    \frac{1}{Z_{\rm ST}^{{\rm NGE}}(\phi)}&\approx& \frac{\sigma_1}{\sigma_N}\frac{\hbar}{\pi \Delta}\frac{L_{J}\cos^{2}{\left(\frac{\phi}{2}\right)}/ [2\cos^2{\left(\phi\right)}]+L_k}{\left(L_{J}/|2\cos{\left(\phi\right)}|+L_k +L_g\right)^{2}}\nonumber\\
    &&+\frac{i}{\omega\left(L_{J}/|2\cos{\left(\phi\right)}|+L_k +L_g\right)}, \label{eq:ZStransmonNGE}
\end{eqnarray}
where $L_J$ is the Josephson inductance of each junction, $L_k \equiv 2L_{kp} + L_{kl}/2$, $L_g \equiv 2L_{gp} + L_{gl}/2$. $L_{kp}, L_{gp}$ ($L_{kl}, L_{gl}$) are the kinetic and geometric inductance of each pad (loop segment), respectively. Note that the pads will dominate the geometric inductance, while the loop dominates the kinetic inductance. In this case, the total inductance is $L = L_{J}/\left|2\cos{\left(\pi\frac{\Phi}{\Phi_0} \right)} \right| + L_k + L_g$.

The flux noise generated due to resident QPs in an NGE split transmon is thus
\begin{eqnarray}
    \tilde{S}_{\Phi}^{\rm NGE \ ST}(\omega) &\approx& \frac{\sigma_1(\omega)}{\sigma_N}\frac{2 \hbar^2 \omega }{\Delta \pi} \left[L_{J}\frac{\cos^{2}{\left(\frac{\pi}{2}\frac{\Phi}{\Phi_0}\right)}}{2{\cos^{2}{\left(\pi\frac{\Phi}{\Phi_0}\right)}}}+ L_k \right] \nonumber \\ &\times& \left[ n_B(\omega) + 1 \right], \label{eq:SphiNGEST}
\end{eqnarray}
when $\cos \left(  \pi\frac{\Phi}{\Phi_0} \right) >0$. When $\cos \left(  \pi\frac{\Phi}{\Phi_0} \right) <0$, replace $\cos^{2}{\left(\frac{\pi}{2}\frac{\Phi}{\Phi_0}\right)}\rightarrow \sin^{2}{\left(\frac{\pi}{2}\frac{\Phi}{\Phi_0}\right)}$ in the numerator multiplying $L_J$ in Eq.~(\ref{eq:SphiNGEST}).

In Eq. (\ref{eq:SphiNGEST}), the term proportional to $L_{J}$ accounts for loss due to QP-tunneling at the junction, while the term proportional to $L_k$ accounts for QP Ohmic loss in the wire. Similarly, in a split transmon with GE junctions, we get
\begin{eqnarray}
    \tilde{S}_{\Phi}^{\rm GE \ ST}(\omega)
    \approx   \frac{\sigma_1(\omega)}{\sigma_N}\frac{2 \hbar^2 \omega }{\Delta \pi} L_k \left[ n_B(\omega) + 1 \right]. \nonumber \\ \label{eq:SphiGEST}
\end{eqnarray}%
Notably, Eqs. (\ref{eq:SphiNGEST}) and (\ref{eq:SphiGEST}) \emph{diverge at zero frequency}.  
This has implications for how \Addition{qubit coherence $|\langle \sigma_+(t)\rangle| \equiv\exp{[-\alpha(t)]}$ decays as a function of time.}

We now calculate the free induction decay (FID) time $T_2^*$ measured in a Ramsey experiment with an ST qubit. From Eq.~(\ref{eq:decayT2}) this is defined as 
$\alpha(T_{2}^{*})=1$, with filter function given by \cite{deSousa2009}
%
%
\beq \label{eq:FIDFilter}
   \mathcal{F}_{\rm FID}(\omega,t) = \frac{1}{2}\left(\frac{\sin(\omega t/2)}{\omega/2}\right)^2.
\eeq
In the limit $t\rightarrow \infty$ finite frequency fluctuations are averaged out, leading to $\mathcal{F}_{\rm FID}(\omega,t) \rightarrow \pi \delta(\omega) t$  and $\alpha(t)$ has a simple linear dependence in $t$. In this limit, $|\langle \sigma_+(t) \rangle|$ follows a simple exponential decay with $\frac{1}{T_{2}^{*}}\propto \tilde{S}_{\Phi}(\omega=0)$. But for QP-induced flux noise, $\tilde{S}_{\Phi}(\omega=0)=\infty$, so finite-frequency fluctuations cannot be averaged out. 

When $t \gg \hbar/(k_BT)$ we can obtain an analytical approximation for $\alpha(t)$. From Eqs. (\ref{eq:decayT2}), (\ref{eq:SphiNGEST}) and (\ref{eq:FIDFilter}),
\begin{eqnarray} \label{eq:FIDalpha}
    \alpha^{\rm ST}(t) 
    &\approx&  x_{\rm QP} \left[\frac{\sin\left(\pi\frac{\Phi}{\Phi_0}\right)}{\cos\left(\pi\frac{\Phi}{\Phi_0}\right)}\right]^2 \left(\frac{\Omega}{\Phi_0}\right)^2  \left(\frac{2 \pi \hbar^2 \Delta}{k_BT}\right)^{1/2} \nonumber \\ 
    &\times&  \left[L_{J}\frac{\cos^{2}{\left(\frac{\pi}{2}\frac{\Phi}{\Phi_0}\right)}}{2{\cos^{2}{\left(\pi\frac{\Phi}{\Phi_0}\right)}}} + L_k \right]  \nonumber \\ &\times&t \left\{ \log \left[4 \left(\frac{k_BT t}{\hbar}\right)^3\right] + 1 - \gamma_E\right\}.
\end{eqnarray}
QP-induced loss thus produces qubit coherence decay $ \propto e^{-t \log(t)}$. 
Again, this is valid for $\cos \left(  \pi\frac{\Phi}{\Phi_0} \right) >0$. When $\cos \left(  \pi\frac{\Phi}{\Phi_0} \right) <0$, replace $\cos^{2}{\left(\frac{\pi}{2}\frac{\Phi}{\Phi_0}\right)}\rightarrow \sin^{2}{\left(\frac{\pi}{2}\frac{\Phi}{\Phi_0}\right)}$ in the numerator multiplying $L_J$. 
Equation~(\ref{eq:FIDalpha}) is valid for both NGE and GE ST; the latter case requires setting $L_{J}=0$.

\begin{figure}
	\centering
	\includegraphics[width=0.45\textwidth]{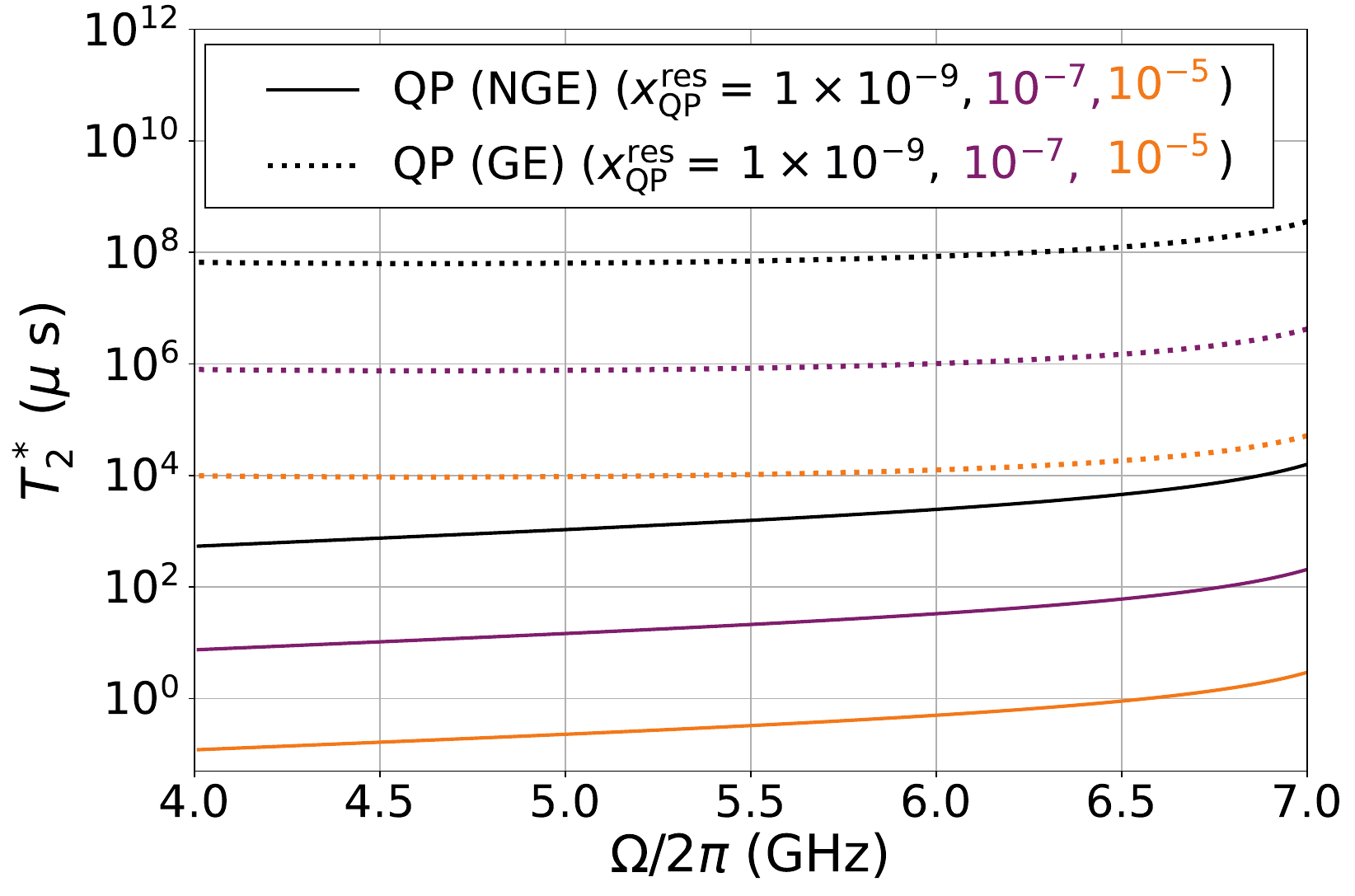}
	\caption{Coherence time $T_2^*$ due to the off-resonant flux noise caused by quasiequilibrium QPs in a split transmon qubit (ST), with $\Omega$ tuned by varying $\Phi/\Phi_0$ from $0.1$ to $0.4$.  
    Solid lines: $T_{2}^{*}$ for a non-gap-engineered (NGE) ST; the $T_{2}^{*}$ is severely limited by flux noise due to QP tunneling across the two junctions. Dotted lines: $T_{2}^{*}$ in a gap-engineered (GE) ST, whose flux noise arises only from Ohmic loss in the ST's loop, leading to a much larger $T_{2}^{*}$. Note these calculations do not include the $1/T_1$ contribution to $1/T_{2}^{*}$.
    Assumed parameters are shown in Table \ref{table1}.} 
\label{fig:T2FID}	
\end{figure}

Figure~\ref{fig:T2FID} shows calculations of $T_{2}^{*}$, obtained by solving $\alpha(T_{2}^{*})=1$ for Eq.~(\ref{eq:FIDalpha}), for both NGE and GE split transmons (Note these calculations do not include the $1/T_1$ contribution to $1/T_{2}^{*}$). 
%
%
We assume a split transmon with the same geometry as sample N1 of \cite{Jaseung2016}. From Eq.~(\ref{eq:Lk}) we estimate $L_k\approx \mu_0 (50~{\rm nm})^{2}\times 1\mu{\rm m}/(0.01\mu {\rm m}^{2}) \times \frac{1}{2}=0.16$~pH,  and assume $E_{J}/E_C = 70$, $L_J =$10 nH. We vary $\Phi/\Phi_0$ from $0.1$ to $0.4$, leading to $\Omega/(2\pi) = 4 - 7$~GHz. 

For NGE STs, we get $T_{2}^{*} =0.1 - 1000$~$\mu$s when $x_{{\rm QP}}=10^{-9}-10^{-5}$. 
The $T_{2}^{*}$ for large QP densities are comparable to values reported in state-of-the-art transmon qubits \cite{Jin2015, Place2021}.  For GE STs, $T_{2}^{*}$ is greatly improved as shown by Eq.~(\ref{eq:FIDalpha}) with $L_{J}=0$. The Ohmic loss in the wires 
leads to $T_{2}^{*}=10^{4}-10^{8}$~$\mu$s for the same range of $x_{{\rm QP}}$.

\section{Conclusions}

In conclusion, we presented a theory of noise and energy relaxation due to resident QPs in superconducting circuits. Our theory uses a generalized impedance method to treat Ohmic loss from QP tunneling across Josephson junctions on the same footing as Ohmic loss in areas of the device away from junctions.


We assumed the resident QPs are in a quasithermal distribution validated by recent experimental observations \cite{Connolly2024} in order to make predictions about energy relaxation and decoherence in several different SC devices. We predicted the impact of resident QPs in CPW resonators, transmons, split-junction transmons, and flux qubits, and compared to other mechanisms such as dielectric loss from TLSs and flux noise from spin impurities. 



We applied our theory to determine the quality factor $Q$ of CPW resonators due to QP Ohmic loss within the superconducting wires (Fig.~\ref{fig:Q_CPW}). The QP contribution to $1/Q$ was shown to be at least $10\times$ smaller than the energy loss due to TLSs in standard dielectric materials. As a result, we conclude that QPs will only be a limiting factor for CPW resonators when the amount of TLSs in the surrounding dielectrics is significantly reduced from current levels. 

We also calculated the contribution of resident QPs to the energy relaxation time $T_1$ in single-Josephson-junction transmon qubits (Fig.~\ref{fig:T1_transmon}). We confirmed that the mechanism of QP tunneling across the junction will dominate $T_1$ in non-gap-engineered (NGE) transmons \cite{Catelani2011}. However, this mechanism is exponentially suppressed in asymmetric gap engineered (GE) transmons. 
For GE transmons, we demonstrate that the dominant QP Ohmic loss arises in the lead electrodes close to the junction, resulting in a reduction of $\frac{1}{T_1}$ by a factor of $2L_k/L_J\sim 10^{-4}$ (here $L_k$ is the kinetic inductance of one of the leads and $L_J$ is the inductance of the Josephson junction). Thus, in GE transmons TLSs will also dominate the energy loss and set the limit on $T_{1}$.


As explained in section \ref{sec:relaxation_noise}, charge noise in an inductive element generate flux fluctuations so it can also be interpreted as flux noise. This effect seems to have been overlooked in the current literature. It is particularly important in cases where flux noise is extracted from $1/T_1$ measurements, as the measured flux noise can originate from three possible sources: Resident QPs, TLSs, and impurity-spin fluctuations.

Our theory shows that charge noise from resident QPs in the SC loop of flux qubits gives rise to a nearly-white flux noise background  (Fig.~\ref{fig:Sphi}). This effect is again dominated by QP tunneling across the junction. 
It provides a possible explanation for the noise background observed in NbTiN devices \cite{Luthi2018, Stavengathesis}, although additional measurements are needed before a definite conclusion is established. 

The magnitude of flux noise due to QP tunneling in NGE junctions is comparable to $1/\omega$ flux noise levels measured in other devices in the $10-1000$~MHz band \cite{Quintana2017}, suggesting the importance of this mechanism as the presence of spin impurities is mitigated \cite{Kumar2016}. Another source of flux noise that we emphasize as important is the one associated to the charge noise of TLSs. 
We find that this can explain the superOhmic flux noise in the GHz range measured in \cite{Quintana2017} (See our Fig.~\ref{fig:Sphi}).

Once again, gap-engineering the junctions can greatly reduce this flux noise background. We show that flux qubits with GE junctions have a significantly smaller flux noise contribution due to QP Ohmic loss away from junctions.

While single-junction transmons are insensitive to off-resonant charge noise \cite{Koch2007}, their nontunable frequency leads to cross-talk due to frequency crowding in quantum computers with a large number of qubits. Circumventing this problem requires qubit designs with flux-tunable $\Omega$, such as the split-junction transmon qubit \cite{Hutchings2017}. Motivated by this, we computed the impact of QP flux noise in the decoherence time $T_{2}^{*}$ of the split-junction transmon (ST) (Fig.~\ref{fig:T2FID}). For NGE STs, our calculation indicates that flux noise from QP-tunneling across the NGE junction might be the limiting factor for $T_2^*$  in current devices. On the other hand, flux noise from QPs is reduced by several orders of magnitude in GE STs.

In conclusion, we described a theory of charge and flux noise due to out-of-equilibrium QP densities in superconducting qubits and resonators. We show that QP tunneling across the junction gives the most important contribution to charge and flux noise. However, this mechanism can be exponentially suppressed when the junctions are gap engineered. In this case we show that the impact of out-of-equilibrium QPs is to produce Ohmic loss and noise in the wires with largest kinetic inductance, such as the small leads connecting the pad electrode to the junction in transmons. Our explicit calculations of Ohmic loss away from the junctions show that the $T_1$ and $T_{2}^{*}$ of GE qubits can be made greater than $10^{5}$~$\mu$s, provided that the amount of TLSs in the dielectrics surrounding the wires can be reduced by several orders of magnitude. Thus, our final conclusion is that asymmetric gap engineering can greatly reduce noise and increase coherence times in superconducting qubits.

\appendix
\section{Fluctuation-Dissipation Theorem for quasiequilibrium distributions \label{sec:GDFT}}


This appendix presents a generalized formulation of the fluctuation-dissipation theorem that shows that it remains valid for quasithermal distributions such as Eq.~(\ref{eq:quasi-eq_dist}).


Assume a system described by a Hamiltonian $\mathcal{H}_0$ is perturbed by an external Hamiltonian $\mathcal{H}_{\rm ext} = - F(t) \hat{O}$ with $\hat{O}$ an observable of interest coupled to its conjugate field $F(t)$. The \emph{susceptibility operator} $\hat{\chi}_O$ is defined according to linear-response theory: $\hat{O}_{F \neq 0}(t) - \hat{O}_{F = 0}(t) = \int_{-\infty}^{\infty}dt' \hat{\chi}_O(t-t') F(t')$. From time-dependent perturbation theory we get $\hat{\chi}_O(t-t') = \frac{i}{\hbar} \theta (t-t') [\hat{O}(t), \hat{O}(t')]$, where $\hat{O}(t)=e^{i{\cal H}_0t/\hbar}Oe^{-i{\cal H}_0t/\hbar}$ is the observable in the interaction picture. In Fourier space
we get $\delta \hat{\tilde{O}}(\omega)=\hat{\tilde{\chi}}_O(\omega)\delta \tilde{{F}}(\omega)$, with susceptibility operator given by
\beq \label{eq:FDT_0}
    \hat{\tilde{\chi}}_O (\omega) = \frac{1}{2 \pi \hbar} \int_{-\infty}^{\infty} d \omega' \frac{[\hat{\tilde{S}}_O(-\omega')]^{\dagger}-\hat{\tilde{S}}_O(\omega')}{\omega - \omega' + i \eta},
\eeq
where $\hat{S}_O (t) = [\hat{O}(t) - \langle \hat{O}(0)\rangle_{\hat{\rho}}][\hat{O}(0) - \langle \hat{O}(0) \rangle_{\hat{\rho}}]$ is the \emph{correlation operator}, $\hat{\tilde{S}}_O(\omega) = \int_{-\infty}^{\infty} d t e^{i \omega t} \hat{S}_O (t)$ is its Fourier transform and $\langle \hat{O}(0) \rangle_{\hat{\rho}} = {\rm Tr} \{ \hat{\rho} \hat{O}(0) \} $ is the average of $\hat{O}(0)$ at a state described by an arbitrary density matrix $\hat{\rho}$. 

Equation~\ref{eq:FDT_0} is an exact operator identity. 
We now specialize to the case where the system is described by a density matrix that is (1) independent of time (i.e. in a steady state) and (2) diagonal in the basis formed by the energy eigenstates $\{ |E\rangle\}$ of ${\cal H}_0$: $ \langle E | \hat{\rho} | E' \rangle = \rho(E) \delta_{E,E'}$ with $\rho(E)$ a real function. Take the average of Eq.~(\ref{eq:FDT_0}) in a such a state and separate its imaginary part to get a \emph{more general version} of the fluctuation-dissipation theorem:
\beq \label{eq:GFDT}
    2 \hbar {\rm Im} \left\{ \langle \hat{\tilde{\chi}}_O(\omega) \rangle_{\rho(E)} \right\}= \langle \hat{\tilde{S}}_O(\omega) \rangle_{\rho(E)} - \langle \hat{\tilde{S}}_O(\omega) \rangle_{\rho(E + \hbar \omega)}.
\eeq
This is based on the identity $\langle [\hat{\tilde{S}}_O(-\omega')]^{\dagger} \rangle_{\rho(E)}= \langle \hat{\tilde{S}}_O(\omega) \rangle_{\rho(E + \hbar \omega)}$, that is valid for a density matrix that satisfies the conditions (1) and (2) above. 
Note that in Eq.~(\ref{eq:GFDT}) the quantities $\langle \hat{\tilde{\chi}}_O(\omega) \rangle_{\rho(E)}$ and $\langle \hat{\tilde{S}}_O(\omega) \rangle_{\rho(E)}$ are the usual susceptibility and power spectral density of $\hat{O}$, respectively; the notation makes their dependence on $\rho(E)$ explicit. 

When $\hat{\rho}$ satisfies the ``quasithermal law'' $\hat{\rho}\propto e^{-{\cal H}/k_BT}$ we get $\rho(E+\hbar\omega)=e^{-\hbar\omega/k_BT}\rho(E)$. 
This implies $\langle \hat{\tilde{S}}_O(\omega) \rangle_{\rho(E + \hbar \omega)}=e^{-\hbar\omega/k_BT}\langle \hat{\tilde{S}}_O(\omega) \rangle_{\rho(E)}$,
leading to the usual fluctuation-dissipation theorem:
\beq \label{eq:FDT}
    \langle\hat{\tilde{S}}_O(\omega) \rangle_{\rho(E)} = 2 \hbar\; {\rm Im} \left\{\langle\hat{\tilde{\chi}}_O(\omega) \rangle_{\rho(E)} \right\}\left[ n_B(\omega) + 1 \right],
\eeq
where $n_B(\omega)$ is the Bose-Einstein distribution. Therefore, the usual fluctuation-dissipation theorem holds for a quasithermal distribution. 


\begin{acknowledgments}
We wish to thank the Engineering Quantum Systems group led by W.D. Oliver at the Massachusetts Institute of Technology for hosting Nava Aquino during part of this work. We acknowledge encouragement and useful discussions with M. Amin, W. A. Coish, M. Hays, P. Kovtun, T. Lanting, W. D. Oliver, K. Serniak, and T. Stavenga.
This work was supported by the Natural Sciences and Engineering Research Council of Canada (NSERC) through its Discovery (RGPIN-2020-04328), and 
its Alliance International Catalyst Quantum Grant (ALLRP-586318-23). 
\end{acknowledgments}

\bibliography{qp_fluxnoise}

\end{document}